# Assessment of physical schemes for WRF model in convection-permitting mode over southern Iberian Peninsula

Feliciano Solano-Farias[1], Matilde García-Valdecasas Ojeda[1,2], David Donaire-Montaño[1], Juan José Rosa-Cánovas[1,2], Yolanda Castro-Díez[1,2], María Jesús Esteban-Parra[1,2], Sonia Raquel Gámiz-Fortis[1,2]

[1] Departamento de Física Aplicada, Universidad de Granada, Granada, Spain

  solanofarias@correo.ugr.es, mgvaldecasas@ugr.es, dadonaire@ugr.es, jjrc@ugr.es, ycastro@ugr.es, esteban@ugr.es, srgamiz@ugr.es

[2]Instituto Interuniversitario de Investigación del Sistema Tierra en Andalucía (IISTA-CEAMA), Granada, Spain

**Corresponding author:** Matilde García-Valdecasas Ojeda (mgvaldecasas@ugr.es)

# Assessment of physical schemes for WRF model in convection-permitting mode over southern Iberian Peninsula

Feliciano Solano-Farias[1], Matilde García-Valdecasas Ojeda[1,2], David Donaire-Montaño[1], Juan José Rosa-Cánovas[1,2], Yolanda Castro-Díez[1,2], María Jesús Esteban-Parra[1,2], Sonia Raquel Gámiz-Fortis[1,2]

[1] Departamento de Física Aplicada, Universidad de Granada, Granada, Spain

[2] Instituto Interuniversitario de Investigación del Sistema Tierra en Andalucía (IISTA-CEAMA), Granada, Spain

## ABSTRACT

Convection-permitting models (CPMs) enable the representation of meteorological variables at horizontal high-resolution spatial scales ($\leq 4$ km), where convection plays a significant role. In this regard, physical schemes need to be evaluated considering factors in the studied region such as orography and climate variability. This study investigates the sensitivity of the Weather Research and Forecasting (WRF) model as CPM to the use of different physics schemes on Andalusia, a complex orography region in the southern part of the Iberian Peninsula (IP). To do that, a set of 1-year WRF simulations was completed based on two "one-way" nested domains: the parent domain (d01) spanning the entire IP with 5 km spatial resolution and the nested domain (d02) for the region of Andalusia at 1 km of spatial resolution. 12 physic schemes were examined from combinations of microphysics (MP) schemes including THOMPSON, WRF single moment 6-class (WSM6), and WRF single moment 7-class (WSM7), and different options for the convection in d01, the Grell 3D (G3), Grell-Freitas (GF), Kain-Fritsch (KF), and deactivated cumulus parameterization (OFF). The simulated precipitation and 2-m temperature for the year 2018, characterized to be a very wet year, were compared with observational datasets from different sources to determine the optimal WRF configuration, including point-to-point and station-point comparisons at different time aggregations (from annual to hourly). In general, greater differences were shown when comparing the results of convection schemes in d01. Simulations completed with GF or OFF presented better performance compared to the reference datasets. Concerning the MP, although THOMPSON showed a better fit in high mountain areas, it generally presents a worse agreement with the reference datasets. In terms of temperature, the results were very similar and, therefore, the selection of the "best" configuration was based mainly on the precipitation results with the WSM7-GF scheme being suitable for Andalusia region.

## 1. Introduction

Nowadays, the notable variations in temperature and precipitation as a consequence of climate change is a topic of great importance to society, given the strong impact that they can cause through highly destructive, short-duration, and localized extreme episodes associated to them. Many studies have widely accepted that global warming associated with the increase in greenhouse gas (GHG) concentrations will intensify the water cycle, increasing global average precipitation and evaporation, with semi-arid areas such as the Mediterranean region suffering some of the most intense impacts (García-Valdecasas Ojeda et al., 2020a, 2020b, 2021a; 2021b; Vicente-Serrano et al., 2017; Sherwood and Fu, 2014). In this framework, regional climate models (RCMs), which have the advantage of describing the regional characteristics such as local topography, have been widely used to project the future climate change at regional scale, promoting initiatives like EURO-CORDEX (Jacob et al., 2014).

In this context, the Weather Research and Forecasting (WRF) model (Skamarock et al., 2021) is an open-source, publicly available code with a set of physically realistic and continuously evolving parameterization schemes. WRF RCM has been already applied over the Iberian Peninsula (IP), a hot spot into the Mediterranean area with high climate variability, showing skill to capture temperature and precipitation changes (Argüeso et al., 2012; García-Valdecasas Ojeda et al., 2020a; Garrido et al., 2020; Marta-Almeida et al., 2016), drought conditions (García-Valdecasas Ojeda et al., 2017) or for producing wind simulations (García-Díez et al., 2015). However, for complex orographic regions, the simulations at spatial resolutions of 10-12 km are still not good enough to capture the precipitation amounts (Takayabu et al., 2022). One of the main sources of uncertainty in RCMs is the parameterization schemes used, particularly those associated with convection (Milovac et al., 2016). This is because convection involves complex physical processes that are not fully resolved by commonly used grid spacings (Déqué et al., 2007). As a result, the use of convection-permitting models (CPMs) has been motivated to address the uncertainties in the representation of clouds, moist convection, and complex topography (Ban et al., 2014). Research has shown that grid sizes of less than 4 km are required for this purpose (Prein et al., 2015) and with the increase in computational capabilities, this task has recently become achievable at the climate scale (Coppola et al., 2020; Lucas-Picher et al., 2021; Pichelli et al., 2021). Recent studies have found added value in models with explicit deep convection compared to their coarser RCM coun-terparts with parameterized convection, especially in the ability of CPMs to represent sub-daily precipitation characteristics in Europe (Ban et al., 2021; Berthou et al., 2020; Caillaud et al., 2021; Fosser et al., 2015; Fumière et al., 2020; Giordani et al., 2023; Lind et al., 2016; Kendon et al., 2017; Leutwyler et al., 2017; Rummukainen, 2016; Torma et al., 2015). However, the dominant dynamics must be considered in the models, so they must be tested to determine the best setup for each region (González-Rojí et al., 2022). Therefore, studies in different regions of the world performed sensitivity simulations of physical parametrizations schemes of WRF (e.g., Di et al., 2019; Gunwani et al., 2020; Tyagi et al., 2018; Xie et al., 2012).

Specifically for the IP, authors such as Argüeso et al. (2011) have addressed this point for horizontal grid spacings ≥ 10 km, showing that climate simulations are sensitive to the chosen parameterization scheme. Generally, the representation of the terrain is more realistic when the resolution is higher, but it is difficult to define the range of convection processes between the subgrid-scale, which can be parameterized by the convective parameterization scheme, and the grid-scale, which must be explicitly resolved (Park et al., 2022). This range commonly is called gray-zone and is between 4-10 km (Hong and Dudhia, 2012). At a grid spacing below 4 km, is generally accepted that a convective parameterization scheme is not required (Borge et al., 2008; Jee and Kim, 2017; Liang et al., 2019), although the simulation of convective processes by explicitly resolved convection could be limited (Arakawa et al., 2016). Meanwhile, there are studies showing that the use for the convective parameterization schemes at these resolutions could improve the simulations (Deng et al., 2006; Lee et al., 2011). Therefore, the convection parameterization schemes validity in the gray-zone must be investigated (Prein et al., 2015).

This study aims to address a comprehensive sensitivity analysis involving different options available in WRF regarding the model physics when representing temperature and precipitation over the southern IP. The paper is structured as follows; Section 2 details the methodology used to configure the sensitivity simulations and the data used in this study; Section 3 describes the results of the analysis, and Section 4 discusses and summarizes the main conclusions.

## 2. Data and methods

### 2.1 The WRF model setup and study area

In the present study, a set of climate simulations was completed using WRF-ARW version 4.3.3 (Skamarock et al., 2021) in convection-permitting mode. The initial and boundary conditions required to run the model were provided by the fifth-generation reanalysis (ERA5, Hersbach et al., 2018) from the European Centre for Medium-Range Weather Forecast, which has a spatial resolution of 0.25° updated at 6-hourly intervals (González-Rojí et al., 2022; Merino et al., 2022; Moya-Álvarez et al., 2018). Following previous studies (e.g., Messmer et al., 2021), WRF has been configured using two one-way nested domains: a coarser domain (d01) with 5 km spatial resolution covering the entire IP, and a finer domain (d02) with 1 km spatial resolution and centered over Andalusia (southern IP, Fig. 1). The domains were configured using a Lambert map projection, with d01 having 320 x 320 grid points in the west-east and south-north directions, and d02 with 576 x 326 grid points. In the vertical, a hybrid vertical coordinate system was defined, with 46 levels up to the top of 50 hPa.

Andalusia, located in southern Spain, between two different water masses (the Mediterranean Sea and the Atlantic Ocean), is a region with a complex topography and extensive coastal region. The Andalusian relief is therefore marked by a significant altitude contrast (Fig. 1), offering a diverse range of ecosystems, such as mountains (Sierra de Cazorla to the northeast, Sierra Morena to the northwest, and Sierra Nevada and Grazalema to the southeast and south, respectively), and valleys, including the Guadalquivir basin that flows through its central part from northeastern to southwestern. This geographical diversity provides a wide range of climates, from subtropical on the Mediterranean coast to continental in the interior.

The WRF simulations were run for a 2-year period in a continuous run, which began on January 1, 2017, and ended on December 31, 2018. According to previous studies (Perez et al., 2022; Zhang et al., 2022), the first year (i.e., 2017) was considered as the spin-up period for all simulations, and, hence, was not included in the sensitivity analysis. Year 2018 was selected according to the Andalusia climatology (considered as the 30-year period between 1992 and 2021), as it was characterized as a wet year. Figure 2 shows the annual precipitation anomalies along with the 2-m maximum and minimum temperature anomalies for the year 2018 compared to the period 1992-2021 for ERA5 reanalysis data. This is consistent with the State Meteorological Agency climate report (AEMET, 2018), which declared that 2018 was the fifth wettest year since 1965 and the second wettest year of the 21st century, only behind 2010.

To select the parameterization schemes, a preliminary analysis was performed for an extreme precipitation event called the Storm Cecilia that occurred between November 21 and 23, 2019. Storm Cecilia was characterized by intense winds, abundant precipitation, elevated waves, and a resulting decrease in temperatures, leading to snow accumulation in the mountainous areas of northern IP, impacting also in France and other countries in western Europe (Gonçalves et al., 2023). For that analysis, 36 configurations were tested as a result of combining different schemes of microphysics (MP), cumulus (CU), radiation (long- and short-wave), and planetary boundary layer (PBL) using the ERA5 reanalysis to force the model every 6 hours and with a 48-h spin-up period. From the model performance of this event (results not shown), we found that the largest changes occurred for the different microphysics and convection of the coarse domain (d01). As a result, PBL and radiation schemes were fixed to those schemes presenting the best results in the experiment. In this regard, the Asymmetric Convective Model version 2 (ACM2; Pleim, 2007) for PBL, and the Community Atmosphere Model 3.0 (CAM3.0; Collins et al., 2004) for both long- and short-wave radiation were selected to further explore the model sensitivity. This selection is also in accordance with previous studies conducted in the study region (Argüeso et al., 2011; García-Valdecasas Ojeda et al., 2017; 2020a). Noah–MP land surface model (Niu et al., 2011; Yang et al., 2011) was used to describe land surface processes.

Among the different options included in the previous analysis, a total of 12 combinations of MP and CU schemes (only for the coarser domain d01) were finally tested and the parameterizations used in each experiment are presented in Table 1. In this regard, the different MP schemes used were: WRF single moment 6-class scheme (WSM6; Hong and Lim 2006), WRF single-moment 7-class (WSM7; Bae et al., 2019) and THOMPSON (Thompson et al., 2008). WSM6 and THOMPSON contemplate ice, snow, and graupel processes and they are recommended for very high-resolution simulations. Otherwise, WSM7 introduces hail as an additional simulated hydrometeor in WSM6. For convection, the parameterizations were Grell-Dévényi three dimensional (G3; Grell and Dévényi, 2002), Grell-Freitas (GF; Grell and Freitas 2014) and Kain-Fritsch (KF; Kain 2004) schemes. Additionally, the coarser do-main was also assessed by disabling the convection scheme (OFF), which allows for the explicit resolution of convection processes.

**Here Table 1** Summary of the twelve combinations of parameterizations used in this study. For convection, the different options for each domain are indicated as well as the acronyms used in this study.

### 2.2 Observational data

As is shown in Table 2, observational datasets from different sources and nature have been employed to analyze the precipitation (pr) and the 2 meters maximum and minimum temperatures (tasmax and tasmin, respectively). This leads to a more comprehensive and accurate interpretation of the data, with potential implications for decision-making related to water and climate management.

**Here Table 2** Observational datasets description and its use in this study. With pr, tasmax, and tasmin, the precipitation, maximum and minimum temperatures are denoted.

The gridded products used in this study were the precipitation from the Climate Prediction Center MORPHing technique (CMORPH; Joyce et al., 2004), developed by the National Oceanic and Atmospheric Administration (NOAA). This provides a quasi-global (60°N - 60°S) precipitation dataset with high spatiotemporal resolution (0.07°) covering the period from 2002 to the present, offering data every thirty minutes. Precipitation at CMORPH is estimated using data from microwave sensors aboard low-orbiting satellites and spatial propagation data derived from infrared data collected by geostationary satellites. The Integrated Multi-satellitE Retrievals version 6 (IMERG; Huffman et al., 2019) Final Run product from the Global Precipitation Measurement (GPM) was also used. IMERG combines information from passive microwave sensors, and infrared sensors to produce high-quality precipitation estimates for quasi-global bands (60° N - 60° S). IMERG, like CMORPH, provides pr values every 30 minutes at a spatial resolution of 0.1° x 0.1°, and covers the period from 2000 to the present as a result of merging ground-based estimates collected in the Tropical Precipitation Measurement Mission (TRMM) satellite during the years 2000 and 2015 with those collected by GPM satellite from 2014 onwards.

Climate Hazards Group Infrared Precipitation with Stations version 2.0 (CHIRPS; Funk et al., 2015) is a high-resolution (0.05º) gridded product that covers the region between 50ºN and 50º S. CHIRPS is the result of a collaboration between the US Geological Survey (USGS) and Earth Resources Observation and Science (EROS) and has a daily temporal resolution spanning the period 1981-present. Unlike traditional methods that rely solely on weather stations, CHIRPS combines data from these stations with satellite-based precipitation estimates provided by National Aeronautics and Space Administration (NASA) and NOAA.

Finally, we also used a high-resolution daily gridded dataset of station-based products from AEMET (AEMET-Grid hereinafter). AEMET-Grid (version 2) provides daily data for tasmax, tasmin, and pr at a spatial resolution of 0.05º x 0.05º from 1951 to 2021 for a grid covering the Peninsular Spain and Balearic Islands (Peral-García et al., 2017).

Although the comparison of model results with multiple sources of information is advantageous; AEMET-Grid data has been selected as the reference database due to its consistency, quality, and high spatial resolution for the study region. AEMET-Grid considers the influence of topography, seasonal variability, spatial distribution, and daily variability of precipitation, as it was generated from data available from 3236 precipitation stations distributed throughout the Spanish national territory. Therefore, it is expected to accurately reproduce the climatology of the region.

Additionally, and as a reference, observations from meteorological stations have been considered at both daily and hourly temporal resolutions, for the three study variables. Fig. 1b shows the spatial distribution of these stations over Andalusia. Note that only stations with less than 10% missing values were included, resulting in a total of 599 precipitation stations across the study region. Of these stations, 303 were provided by AEMET, 4 by the Autonomous Agency for National Parks (OAPN), 119 were stations from the Automatic Hydrological Information System (SAIH) from Hidrosur network (SAIH-S) and 173 from Guadalquivir Basin network (SAIH-G). For temperature, a total of 219 stations were utilized including tasmax and tasmin data, with 184 stations from AEMET, 6 from OAPN, and 29 from SAIH-G (see Table 3).

**Here Table 3** Meteorological stations used in this study. Different sources were used: (1) the Spanish Meteorological Agency (AEMET), (2) the Autonomous Agency for National Parks (OAPNs), and the Automatic Hydrological Information System (SAIH) from (3) Hidrosur network (SAIH-S) and (4) from Guadalquivir Basin network (SAIH-G).

### 2.3 Comparison methods

Several analyses were performed to determine the best combination of parameterization schemes to characterize the climate conditions in Andalusia using WRF. For the point-to-point comparison, every gridded dataset was remapped according to the AEMET-Grid coordinates using a bilinear interpolation method. In the comparison with stations, we used the nearest grid point of the corresponding dataset using the nearest neighbor method. Different time aggregations were considered (from annual to daily time scales) and were made to identify differences in model characterization when different parameterization schemes were used. The analysis focused on comparing the WRF outputs with reference datasets at annual scale, but also the comparison was made for autumn season (covering September-October-November, SON) because this is the season when more convective precipitation is expected in the study area.

In addition, considering the complex orography of the region (Fig. 1b), the annual cycles of the monthly precipitation and temperature were analyzed for different ranges of altitude to further investigate the effect of the orography in the different configurations. Thus, the monthly average was first calculated for each grid point, and then three regions based on altitude within Andalusia were distinguished to compute their spatial average (Table 4): (1) low altitudes (LAs) that cover regions lower than 400 m (1291 grid points); (2) middle altitudes (MAs), from 400 to 1500 m (1456 grid points); and (3) high altitudes (HAs), for regions higher than 1500 m (85 grid points). On the one hand, LAs contains most of the stations, 315 stations for pr (52.58%) and 121 stations for tasmax and tasmin (55.25%). On the other hand, for MAs there are 276 pr stations (46.08%) and 94 temperature stations (42.92%). The remaining region, HAs, is characterized as the smallest area (mainly Sierra de Cazorla and Sierra Nevada), with only 8 pr stations (1.33%) and 6 temperature stations (1.83%).

**Here Table 4** Regions used in the study of the annual cycles of the monthly precipitation and temperature.

For daily pr, tasmax and tasmin values, WRF outputs were compared with the reference datasets using the Kling-Gupta Efficiency (KGE; Gupta et al., 2009; Kling et al., 2012). KGE is a goodness-of-fit measure widely used to assess the ability of hydrological models (e.g., García-Valdecasas Ojeda et al., 2022; Hafizi and Sorman, 2022). However, it can also be used to analyze the performance of regional models (e.g., Beck et al., 2019, Merino et al. 2021). The KGE considers different types of model errors, namely the error in the mean, the variability, and the dynamics. Eq. (1) details KGE computing:

$$KGE = 1 - \sqrt{(\beta - 1)^2 + (\alpha - 1)^2 + (TSC - 1)^2} \quad (1)$$

where $\beta$ represents the mean bias between observations and simulations, $\alpha$ measures the variability, and TSC the temporal correlation. In this study, a modified version of KGE proposed by Pool et al. (2018) was used, which considers variability and correlation by using nonparametric components. For this metric, TSC is measured using the Spearman correlation, and $\beta$ is computed with Eq. (2).

$$\beta = \frac{\overline{x}_{sim}}{\overline{x}_{obs}} \quad (2)$$

To compute $\alpha$, flow duration curves (FDCs) were calculated. That is, for each grid point, time series (x) of both simulations and observations, were sorted in descending order to obtain x' timeseries following Eq. (3):

$$x' = sort(x) \quad (3)$$

The FDCs is thus the result of applying Eq. (4)

$$FDC = \frac{x'}{n\overline{x}} \quad (4)$$

where n and $\overline{x}$ represent the number of records and the mean in each timeseries respectively. Finally, $\alpha$ is computed as:

$$\alpha = 1 - \frac{1}{2}\sum_{i=1}^{n}|\text{FDC}_{\text{sim}}(i)\text{-FDC}_{\text{obs}}(i)| \qquad (5)$$

where $\text{FDC}_{\text{sim}}(i)$ and $\text{FDC}_{\text{obs}}(i)$ represents FDCs for simulations and observations, respectively.

In order to investigate the capability of WRF to capture extreme values, different percentiles of precipitation on wet days (pr > 1 mm), maximum, and minimum temperatures were analyzed for each region defined for the calculation of the annual cycles (i.e., LAs, MAs, and HAs). Thus, the 50th, 60th, 70th, 75th, 80th, 90th, 95th, and 99th percentiles of each of the variables were calculated for each WRF configuration as well as the AEMET-Grid values.

Finally, two different extreme precipitation events that occurred in Andalusia in 2018 were examined, each with a different nature and both of particular interest to this region. The objective of this latter analysis was to determine if WRF is able to reproduce the spatial patterns as well as the intensity of the events and to elucidate which combination of parameterizations best represented them. The first event selected was Storm Emma. This low-pressure system impacted the IP from February 28th to March 4th and it was one of the most intense and widespread episodes in the 2018 year. The second extreme event was a mesoscale convective system (MCS) that occurred between October 20 and 21, resulting in rainfall of more than 300 mm per day in some areas of Andalusia.

## 3. Results

### 3.1 Total precipitation and mean temperatures

The WRF performance when using different parameterizations is evaluated by analyzing the spatial distribution of annual accumulated precipitation over the entire region of Andalusia and for autumn (Fig. 3). As for annual values (Fig. 3a), in the first row, observational data (i.e., AEMET, CHIRPS, CMORPH, and IMERG) are shown, while the results obtained from the 12 WRF simulations are displayed in the remaining three rows. Dots in the maps represent the annual precipitation from observational stations. The $r_G$ is used to denote the pattern correlation (i.e., the Spearman spatial correlation) between AEMET-Grid and the corresponding dataset (i.e., observational datasets or WRF outputs) and $r_S$ represents the pattern correlation between stations and the nearest grid point in each case. The mean absolute error ($\text{MAE}_G$ and $\text{MAE}_S$ for AEMET-Grid and stations, respectively) is also displayed. Both metrics are calculated in order to elucidate the combination of parameterizations that outperform the others over the entire territory.

As can be seen in Fig. 3a, the annual precipitation in 2018 presents some interesting results. IMERG, in general, shows a higher amount of precipitation, which is closer to the reference datasets ($MAE_G = 138.78$ mm and $MAE_S = 160.47$ mm) than both CHIRPS ($MAE_G = 207.10$ mm and $MAE_S = 180.39$ mm) and CMORPH ($MAE_G = 221.29$ mm and $MAE_S = 217.84$ mm). However, for pattern correlations, CHIRPS ($r_G = 0.77$ and $r_S = 0.64$) and CMORPH ($r_G = 0.61$ and $r_S = 0.49$) show higher values than IMERG ($r_G = 0.56$ and $r_S = 0.51$). This suggests that CHIRPS and CMORPH, despite displaying lower total precipitation, provide a more accurate representation of the pr spatial distribution in this region, at least for the year 2018.

Regarding the comparison between WRF and the reference datasets, the results show that the model is able to capture reasonably well the spatial patterns of precipitation in Andalusia (pattern correlations up to 0.81 and 0.73 in relation to AEMET-Grid and stations, respectively). That is, WRF is able to represent the annual pr as well as or even better than observational products when comparing to AEMET-Grid. Thus, two groups of parametrization configurations can be identified based on their agreement with AEMET-Grid and stations. The first group, consisting of the simulations with G3 and KF CU schemes, shows lower pattern correlation ($r_G < 0.78$ and $r_S < 0.7$) and higher MAE ($MAE_G > 220$ mm and $MAE_S > 190$ mm). In contrast, those combinations using GF and OFF CU, exhibit higher pattern correlation ($r_G > 0.78$ and $r_S > 0.7$) and lower MAE ($MAE_G < 209$ mm and $MAE_S < 181$ mm). When MP schemes are compared, however, small differences are observed between the different cumulus schemes, with THOMPSON microphysics producing, in general, lower precipitation than those that use WSM6 and WSM7. Therefore, the best performance of WRF seems to be associated with those simulations using the GF and OFF CU. The latter is especially true for simulations completed with WSM7 with $MAE_G$ values lower than 180 mm. However, significant discrepancies are observed when G3 and KF are used, with the WRF underestimating the observed precipitation in general.

The results also show that all WRF simulations tend to overestimate the precipitation in the mountainous regions of Sierra Nevada and Sierra de Cazorla. Conversely, the model shows an underestima-tion of precipitation over low altitude (< 400 m) regions in the western and central part (e.g., Guadalquivir Basin) and eastern part of Andalusia.

Similar conclusions can be drawn for the autumn accumulated pr (Fig. 3b). In this case, WRF has a lower agreement with AEMET-Grid and stations in terms of pattern correlations (pattern correlations up to 0.69). WRF simulations with the KF convection scheme show the lowest spatial pattern ($0.37 < r_G <$ 0.49, and $0.34 < r_S < 0.44$) and maximum error values (85 mm $< MAE_G <$ 108 mm), while those completed with G3, GF, and OFF CU present a better performance ($0.54 < r_G$ and $r_S <$ 0.69 and 57 mm $< MAE_G$ and $MAE_S <$ 77 mm). THOMPSON-GF, THOMPSON-OFF, and WSM7-GF appear to outperform other combinations, with the latter presenting the highest pattern correlation with stations ($r_S = 0.64$) and the lowest error ($MAE_S = 58.85$ mm). As for annual values, GF and OFF CU schemes along with WSM7-MP, show the highest r and the lowest MAE values. Concerning the other observational data for autumn, IMERG has a spatial pattern closer to AEMET-Grid, with pattern correlation of 0.80, being the correlation with stations also acceptable ($r_S = 0.68$).

Fig. 4 shows the annual tasmax and tasmin for WRF simulations compared with AEMET-Grid. Additionally, the mean temperature values from stations are shown with dots. Results for autumn temperatures are shown in Fig. S1 (supplementary material). In general, the results for tasmax and tasmin, in both annual and autumn scale, are consistent with those from precipitation. WRF, performs satisfactorily the spatial pattern of both temperatures when comparing observational data, with the most pronounced biases in the mountainous areas of Sierra Nevada and Cazorla. In terms of tasmax (Fig. 4a), high pattern correlations are observed between WRF and stations. For all WRF combinations, $r_S$ takes values up to 0.83 and $r_G$ is around 0.94. However, MAE values indicate low differences: 0.73ºC $<$ $MAE_G <$ 0.82ºC and 0.91ºC $< MAE_S <$ 0.98ºC. Concerning the tasmin (Fig. 4b), the correlation slightly decreases, with $r_G$ values between 0.88 and 0.89 and $r_S \sim 0.77$, while MAE values slightly improve: 0.69ºC $< MAE_G <$ 0.75ºC and 0.92ºC $< MAE_S <$ 0.93ºC. These results show the ability of WRF for simulating daily maximum and minimum temperature in this region, with all configurations showing very similar results.

### 3.2. Annual cycle of monthly precipitation and temperature

Fig. 5 shows the annual cycle of monthly pr, tasmax, and tasmin (rows), from AEMET-Grid, stations, and the outputs from the 12 WRF simulations for the three ranges of altitude (i.e., LAs, MAs, HAs, columns). In general, annual cycles from WRF are comparable to AEMET-Grid and stations. However, for pr in the LAs region (Fig. 5a), WRF consistently underestimates pr by nearly all configurations. This is particularly evident for THOMPSON-G3 and THOMPSON-KF, which show a difference of around 130 mm in March, the wettest month. In contrast, THOMPSON-GF and combinations with deactivated convection (OFF) produce the best results for LAs. However, WRF overestimates the pr in HAs (Fig. 5c), where THOMPSON-G3 and THOMPSON-KF exhibit the lowest overestimation. For these elevations, the combinations using WSM6 and WSM7 tend to overestimate by approximately 160 mm when simulations are compared to both AEMET-Grid and stations. For MAs (Fig. 5b), WRF presents the best performance for the annual cycle with a near perfect fit when THOMPSON-GF, THOMPSON-OFF, and WSM7-OFF (WSM6-OFF) are used in relation to stations (AEMET-Grid). During autumn (SON), a secondary peak in pr is recorded, as is illustrated in the first row of Fig. 5. For this season, the model configurations suggest a minor improvement when KF is used. Conversely, more satisfactory results are achieved when employing combinations of WSM6 and WSM7 with G3, GF, and OFF. For tasmax (Fig. 5d-f) and tasmin (Fig. 5g-i), all configurations provide very similar results. The better fits related to AEMET-Grid and stations are observed for LAs and MAs regions. This can be attributed to the uniform topography of the region under study (central Andalusia), which allows temperature variations to be better captured. In contrast, in regions with more complex topography, such as HAs, WRF tends to exhibit larger discrepancies (Figs. 5f and 5i), especially for tasmax (Fig. 5f). Here, the simulations consistently underestimate tasmax by +1.0°C to +3.5°C compared to stations and AEMET-Grid, respectively.

### 3.3. Daily values

The results of the KGE metric are depicted in Fig. 6 for annual and autumn pr, comparing the different WRF configurations with both AEMET-Grid and stations. KGE ranges from minus infinite to 1, where 1 indicates a perfect fit and values below -0.41 mean a poor model performance (Knoben et al., 2019). At an annual scale (Fig. 6a), KGE values for stations with respect to AEMET-Grid show quite good results, as expected because AEMET-Grid data is obtained from AEMET station observations.

However, the other observational gridded products show low KGE values for most of Andalusia, except IMERG, which presents values above 0.6 in a large part of the region. For WRF simulations, however, higher KGE are shown, at least for combinations using GF and OFF CU, both with respect to AEMET-Grid and stations data. In a large part of the study region, both combinations have KGE values greater than 0.52, especially in the central east. It is worth mentioning that the combinations of WSM7 MP with GF and OFF CU consistently produce higher KGE (values greater than 0.56 with respect to AEMET-Grid and 0.40 with stations), across almost the entire Andalusian region. Contrarily, a worse performance is shown for G3 and KF, with KGE values up to 0.40 with respect to AEMET-Grid and stations, and KGE values close to 0 in the northwestern region and Grazalema when compared to stations data. Sierra Nevada has the lowest agreement with AEMET-Grid (KGE values less than -0.41). However, it is important to consider that observational gridded products in mountain regions could be more affected by measurement errors due to a lack of stations.

As can be seen from the analysis of the different KGE parameters, TSC (Fig. S2a in supplementary material) for CHIRPS and CMORPH is lower than for the WRF simulations, especially in central regions where low KGE values seem to be due to a lower TSC. In addition, all WRF simulations show high TSC values in high mountain regions, such as Sierra Nevada, suggesting that low KGE values are not due to this parameter. GF and OFF CU exhibit slightly higher TSC values than other configurations, in general, especially when they are combined with THOMPSON MP. In the same way, the better performance in terms of variability ($\alpha$, Fig. S2b in supplementary material) is shown for the pr simulated with GF and OFF CU ($\alpha$ values higher than 0.8 in much of Andalusia). For this parameter, the results also show that IMERG presents substantially greater values. Concerning the mean bias ($\beta$, Fig. S2c in supplementary material), the results show that both observational gridded products and simulations are generally underestimated when compared to AEMET-Grid and stations except for IMERG. However, mountainous regions show a great overestimation in relation to this parameter, explaining the low KGE values in these grid-points. For IMERG, although $\beta$ values are closer to one in many areas, there are both overestimations and underestimations. Therefore, for the latter dataset, high KGE values seem to be the result of the high agreement between IMERG and AEMET-Grid (and stations) in terms of $\alpha$.

KGE results for autumn (Fig. 6b), show a similar behavior to annual scale, although with worse adjustment. However, for this season, KGE values obtained by G3 convection are found to be comparable to those generated by GF and OFF, with values above 0.40 across most of the region. In contrast, KF exhibits considerably lower KGE values in several zones, with relatively large areas showing values below 0.20. On the other hand, in Sierra Nevada, consistent values below -0.40 are observed across all WRF configurations, except for THOMPSON MP. The analysis of the KGE components for autumn pr (Fig. S3 in supplementary material) shows results in line with the corresponding annual case. The highest TSC values compared to AEMET-Grid and stations are shown for IMERG and for WRF simulations performed with G3 and especially with GF and OFF CU (Fig. S3a, in supplementary material). When comparing MP schemes, however, there are slightly higher values for WSM6 and WSM7. In terms of variability (Fig. S3b), simulations with GF appear to outperform the other configurations, reaching values closer to 1 in some regions over the east. Concerning mean bias, Fig. S3c, all configurations tend to underestimate in a large part of Andalusia, although the bias patterns are not as homogeneous as at annual scale. WRF seems to present higher bias than for the annual case, but it still tends to underestimate pr in western and central areas, while overestimations appear in the eastern part. Again, slightly better results are obtained with G3, GF, and OFF convection schemes.

For annual temperature, Fig. 7 depicts the results of the KGE metric for tasmax and tasmin. In alignment with previous findings, WRF shows a good ability to reproduce tasmax and tasmin across all configurations in relation to observational data. The KGE values for tasmax (Fig. 7a) are marked by high positive values ($> 0.86$) across nearly the entire region, both for AEMET-Grid and stations. However, greater discrepancies are presented over the highest elevations ($> 1500$ m), where KGE values range from 0.50 to 0.80. Consequently, it is difficult to discern at first glance which configuration provides the optimal fit. Nevertheless, when analyzing each component of the KGE separately for tasmax, TSC shows values closer to 1 (Fig. S4a, in supplementary material), with the lowest correlation for coastal regions. For this KGE terms, all simulations seem to present almost identical values. Similarly, the variability of tasmax is also realistically reproduced, with α values exceeding 0.96 except for Sierra Nevada and Sierra de Cazorla (Fig. S4b in supplementary material). Referring to β (Fig. S4c), all simulations present a generalized underestimation in relation to reference data. The best fit is found in the Guadalquivir basin, where tasmax is less underestimated, while the highest underestimations are found in the easternmost part of the region.

For tasmin (Fig. 7b), a slight decrease in the KGE is observed, which generally ranges from 0.80 to 0.90, with the Guadalquivir basin presenting areas with a KGE greater than 0.90. Moreover, as for tasmax, the lowest KGE values are found in high-altitude regions, where the values fall below 0.50. When the KGE parameters are analyzed separately, it can be seen that all configurations show very similar values. In terms of TSC (Fig. S5a), the values are in a very similar range to tasmax. However, the spatial patterns of tasmax are very different from those found for this variable. In addition, the KGE patterns seem to be more influenced by $\alpha$ and $\beta$. In terms of $\alpha$ (Fig. S5b), the results show that the WRF simulations better capture the variability of tasmin in western Andalusia with values close to 1. In terms of $\beta$ (Fig. S5C), and unlike tasmax, tasmin show both overestimations and underestimations.

Fig. 8 shows the q-q plot calculated for wet-day pr (a-c), tasmax (d-f), and tasmin (g-i) for observations from the AEMET-Grid product versus the different WRF configurations for each specified region based on altitude (i.e., LAs, MAs, and HAs). The percentiles from station observations are also displayed in Fig. 8 for comparison. In the panels, the gray line represents a perfect fit between simulations and AEMET-Grid. Values below this line indicate WRF underestimation in comparison to AEMET-Grid, while values above this perfect fit line indicate WRF overestimations. For LAs (Fig. 8a), all configurations are very similar except for the highest percentiles where the model overestimates the precipitation, with the WSM6-OFF showing the greatest overestimation when compared to AEMET-Grid. This overestimation, however, presents a better fit with AEMET stations. Similar behavior appears to occur for MAs (Fig. 8b), though the results for the highest percentile appear to show greater differences. For HAs, WRF seems to have greater difficulty to represent this variable as is evidenced by a greater discrepancy with AEMET-Grid. For temperature, both tasmax and tasmin are adequately represented by all WRF configurations with a very similar behavior and a better fit than with the stations, especially for the case of tasmin in HAs (Fig. 8i).

Finally, an analysis of extreme precipitation events with different physical nature has been carried out in order to explore the ability of WRF simulations under different parameterization schemes to capture extreme rainfall events. For this study, firstly, the Storm Emma event was analyzed. This event was composed by a deep depression formed on February 24 and dissipated on March 5, with a severe impact on a large part of Western Europe (Gonçalves et al., 2023). According to AEMET, Storm Emma was formed to the southeast of Newfoundland, and it rapidly deepened during the 25[th] as it moved southeastward toward the Azores Islands, while a deep occluded front was formed. The low-pressure system began to move northwards, affecting the IP, on February 28 (Fig. 9a-d). During this period, Storm Emma brought heavy rainfall to this region, with up to 120 mm of rain falling in some areas of Andalusia. On March 1, the low-pressure center was located to the north of Spain, and its movement became slower over the next 24 hours, remaining nearly stationary on March 3 and 4. During this time, when up to 200 mm/day falling in south of Andalusia, Storm Emma was absorbed by a later low-pressure system, which was shallower and less active.

Fig. 10 shows the accumulated precipitation recorded from 28[th] February to 4[th] March by both observations and WRF simulations. According to AEMET-Grid, the highest amounts of precipitation fell in the Grazalema, with accumulated values above 300 mm during Storm Emma. These values are also reproduced, at least in part, by CHIRPS, CMORPH and IMERG. Concerning the results from the WRF simulations, it can be seen that G3 and KF CU underestimate, in general, the total amount of precipitation across the region, whereas GF and OFF underestimate the rainiest areas and overestimate the western and central regions. In general, GF and OFF generate a slightly better fit than G3 and KF CU configurations. The best fits for the whole region seem to be for THOMPSON-OFF, THOMPSON-WSM6 and WSM7-GF, with this latter giving the most similar accumulated pr values to AEMET-Grid in the regions of maximum precipitation. Furthermore, all WRF simulations accurately capture even the temporal evolution of this episode, although the total amount of precipitation varies between configurations. WSM6 and WSM7 MP with GF and OFF CU perform better at representing the temporal evolution of precipitation (Fig. 9e-f).

The second extreme event examined was the heavy precipitation event that occurred in autumn, between October 20 and 21. This precipitation event was caused by two convective systems (Figure 11a-b), which resulted in heavy rain, primarily in southern Andalusia. Fig. 11c-d displays the accumulated precipitation from 20[th] to 21[st] October from observations and WRF simulations. Precipitation exceeding 150 mm fell in southern Andalusia during this two-day event, as shown in AEMET-Grid, which was also captured by CMORPH and to a lesser extent by IMERG. CMORPH and IMERG show values in a more dispersed region, whereas AEMET-Grid shows precipitation for a more localized region (black box in Fig. 12). In general, the results show that WRF presents a greater difficulty in capturing the spatial pattern of this event, probably because it is a very localized extreme event in time and space. In this sense, it should be remembered that the simulations are not intended to capture a punctual event that occurred 1 year and almost 11 months after the beginning of the simulation. In a comparison between parameterization schemes, the results show that in general KF is not able to reproduce the event while the rest of CU shows a better performance for the event. Among all the configurations, WSM7-GF is the one with the best characterization of the event.

## 4. Discussion and conclusions

The aim of this study was to determine the best configuration to complete long-term simulations with the Weather Research and Forecasting model in convection-permitting mode for Andalusia, a region with a complex topography located in southern IP. Different physics schemes were used to generate simulated precipitation and temperature at very high spatial resolution (1 km) for the year 2018, which was considered as a wet year. The sensitivity analysis was based on 12 1-year climate simulations that resulted from combining two types of parameterizations (convection and microphysics), which are expected to have a greater impact on convection-permitting simulations. A previous analysis that included additional combinations fixed the radiation, land surface, and boundary layer schemes. Thus, the Grell-Freitas, Grell 3 dimensional, Kain-Fritch, and OFF parameterizations were combined for the coarser domain, d01. The latter, known as convection deactivation, was used to investigate the impact of explicitly resolving convection at a spatial resolution within the so-called gray zone. In this case, the THOMPSON, WSM6 and WSM7 schemes were chosen for microphysics.

Overall, WRF demonstrated a good ability to capture the spatial patterns of the annual accumulated precipitation over Andalusia. Our results showed that the model capability was even better than other observational gridded products such as CHIRPS, CMORPH and IMERG when compared to the reference datasets (AEMET-Grid and stations). In the same context, authors such as Wagner et al. (2018) found that WRF run at high spatial resolution (circa 1 km) outperformed satellite products over complex geographical regions in Germany. Our findings also revealed that WRF had more difficulty representing precipitation spatial patterns during autumn. The latter could be attributed to the occurrence of more convective events during this season (Argüeso et al., 2011; Casas-Castillo et al., 2022; Esteban-Parra et al., 1998; Hidalgo-Muñoz et al., 2015), which are more difficult to detect because they typically occur very locally and suddenly.

In general, the results here found also evidenced the high uncertainty associated with precipitation performance based on observations from different sources. These were even greater than the uncertainty associated with the use of different WRF configurations in terms of microphysics and convection. Moreover, the results of the comparison between configurations revealed that convection parameterization over the coarser domain, d01, had a greater impact on simulations than microphysics, as previously suggested Yang et al. (2021). Thus, WRF simulations completed with GF or with convection in d01 turned off demonstrated better agreement with observations, in general. In contrast, simulations carried out with KF and G3 resulted in greater underestimation of the accumulated pr, with KF exhibiting the lowest pattern correlation in all cases. These results agree with previous studies that found inadequate performance running WRF using KF in areas such as the Euro-CORDEX domain (Mooney et al., 2013) and the IP (García-Valdecasas Ojeda, 2018). Although both studies were conducted at coarser resolutions, both, like our study, found high bias and generally low temporal correlations with observations for simulations with KF. Moreover, our findings were consistent with those from Wagner et al. (2018), who concluded that GF and OFF convection provide similar results in representing pr at high resolution scales ($\leq$ 5 km) over a complex topography. This appears to be due to the nature of GF parameterization. That is, at a spatial resolution of 5 km, GF is rather inactive (Grell and Freitas, 2014).

In terms of the annual cycle and extreme values, WRF also showed comparable results to those from AEMET-Grid and stations. However, the ability of the model varied depending on the region. The model tended to underestimate the precipitation in regions with lower-altitude, particularly when THOMPSON-G3 and THOMPSON-KF were used. For middle altitude areas, WSM6-OFF appeared to be the best configuration followed by WSM6-GF and WSM7-OFF. Moreover, it was observed that the best MP parameterization corresponded to WSM7 when comparing the precipitation results for the entire region. These results agree with Borge et al. (2008), who found that WSM6 was the best MP parameterization for simulations in a 3 km domain over the IP. THOMPSON-G3 and THOMPSON-KF, however, showed the best fit at high-altitude. These findings are consistent, at least in part, with those of Karki et al. (2018) and Varga and Breuer (2020), who discovered that the THOMPSON MP was better at capturing accumulated precipitation in high altitude regions (Central Himalaya and Carpathian Basin, respectively). Furthermore, when convection was disabled, the model tended to overestimate precipitation in high-altitude regions, which is consistent with the findings of Messmer et al. (2021). However, it is important to note that the number of stations and grid points in this region is quite limited, making it difficult to draw firm conclusions about the accuracy of these results. In this regard, Merino et al. (2021) in an evaluation of the impact of different factors on the development of precipitation gridded datasets pointed out the station density as a critical factor for the development of gridded datasets. Similarly, Hofstra et al. (2010) highlighted smoothed behavior in a grid when the number of stations for the interpolation is limited. Similar conclusions can be drawn from the KGE analysis as WRF demonstrated a good agreement with respect to the reference datasets except for certain areas in Sierra Nevada. For this region, KGE values below -0.41 were found. This result was expected, however, given that Sierra Nevada is a high mountain region with high spatiotemporal precipitation variability, a higher fraction of solid precipitation, and a small number of stations (Prein and Gobiet, 2017). All this implies that the representation of precipitation in gridded products for this area may be insufficient, with large discrepancies between databases, as was seen between CHIRPS and other observational products.

For temperature, WRF demonstrated good agreement with observations, especially when the maximum temperature was considered. However, when we compared the different configurations with the purpose of selecting the best to capture the climate on this region, we found that all simulations showed similar spatial patterns, thus not being a determinant factor for the choice of the best WRF configuration. Similar results were found in Argüeso et al. (2011), who completed a sensitivity analysis for our study region using WRF and in terms of precipitation and temperature. In another region, Somos-Valenzuela and Manquehual-Cheuque (2020) in a study to identify a set of parameterizations to characterize the Patagonian climate with WRF, found a very similar behavior for all configurations in terms of temperature. Therefore, temperature does not appear as a conclusive variable when selecting the model configuration.

The model ability to capture extreme precipitation events occurred in 2018 was also analyzed. Note that this study focused on analyzing long-term regional simulations, and not on capturing specific events. In fact, after more than a year of continuous simulation, the model has developed its own dynamics. However, other aspects could also influence the representation of these events, especially for the October MCS. For example, precipitation is very sensitive to the PBL parameterization (e.g., Moya-Álvarez et al., 2020; Merino et al., 2022), so for the representation of these events it might be appropriate to use a different scheme than the one chosen for our climate-mode simulations. In this case, we set the PBL scheme to ACM2 according to previous studies (e.g., Argüeso et al., 2011), that tested different combinations of parameterization using climate-mode simulations in our study region. However, this does not mean that ACM2 is the most appropriate for all precipitation events. Moreover, we could improve the model's characterization of these events by feeding the model with boundary conditions at a higher frequency. In climate mode, however, this aspect is a computational challenge, and the purpose is not so much to represent specific events, but an average behavior of them. The use of spectral nudging could also be beneficial in convection-permitting mode. However, the impact of this methodology on simulations that allow for convection needs further investigation (Prein et al., 2015), and if we want to compare our results with other studies, it is better not to use it. In this regard, Dominguez et al. (2023) and Kukulies et al. (2023), found that the differences between simulations conducted with and without spectral nudging were negligible. In any case, it can be concluded that WRF presented a high capability to represent this type of event. In addition, we found that WRF represented better the Storm Emma than the October MCS, showing in the first case a temporal evolution very similar to observations. This could be related to the greater ability of the model to represent dynamic structures at synoptic scales (Guo et al., 2020), as in the case of Storm Emma. Furthermore, Yang et al. (2019) proposed that WRF has a significantly higher accuracy in representing even rain than uneven rain. Moreover, for this analysis, the results also showed that microphysics parameterization played a more significant role in representing precipitation events than convection parameterization. In this regard WSM6, and especially WSM7, were the best MP configurations for representing these extreme events.

Although this study does not aim to evaluate the performance of satellite-based observation products, it would be interesting to highlight the main conclusions drawn from the comparison of these datasets.

In general, IMERG showed more similar values to AEMET-Grid than CHIRPS and CMORPH in both annual (accumulated precipitation) and daily (i.e., KGE) values. However, the latter were able to better reproduce the spatial patterns of precipitation. Compared to the weather stations, CHIRPS, CMORPH, and IMERG underestimated accumulated precipitation during Storm Emma and the October MCS, while IMERG accurately reproduced both accumulated precipitation and its spatial distribution, at least for Storm Emma. Our results agree at least in part with those of Navarro et al. (2019), who found that the IMERG accurately reflects the patterns of European precipitation. Similarly, Navarro et al. (2020) pointed out that the IMERG precipitation product is a good option to represent the precipitation when station coverage is poor in areas of complex topography such as the Ebro Basin. For the Iberian Peninsula, moreover, Tapiador et al. (2020), emphasized that IMERG is able to describe local precipitation patterns and is reliable as an additional source of information to the official AEMET rain gauge network.

Overall, this study is the baseline to determine an optimal setup for WRF to correctly simulate the observed precipitation patterns and amounts in Andalusia, a very complex region in southern IP. This is also the first time that this region was resolved at a convection-permitting spatial resolution of 1 km, so several configurations, combining microphysics and cumulus, have been tested in the WRF model. Our results, in general, evidenced that WRF is a very valuable tool to be used to carry out climate simulations in convection-permitting mode in the region of Andalusia in terms of both precipitation and temperature. Moreover, our results for simulations in d01 domain suggested that simulations with convection deactivated in the gray zone (from 4 to 10 km) appear to better capture the observed precipitation patterns than those with convection parameterized. Only simulations with d01 employing GF CU exhibited a similar behavior. The latter is likely due to GF CU, as previously discussed, is rather inactive at the spatial resolution of 5 km. Other studies also found a better fit when the parameterization was deactivated in the gray-zone (e.g., Singh et al., 2018). Specifically, Douluri and Chakraborty (2021) demonstrated that the deactivated convection parameterization generates a benefit compared to G3 and KF, in line with our results. Therefore, we can finally conclude from these results that, while no combination appears to outperform the others for all regions, variables, and analysis, as occurred at coarser resolution according to Argüeso et al. (2011), WSM7-GF appears to be a better option for this domain. These find-ings are crucial for adequately simulating the long-term climate over Andalusia based on temperature and precipitation at very-high spatial resolution, in order to generate climate change projections and thus, to improve our knowledge of the potential impacts of climate change in this region.

## Acknowledgements

The authors would like to thank the anonymous reviewers for their help in improving the work. F. Solano Farías acknowledges the Mexican National Defense Secretary for the predoctoral fellowship. This research has been carried out in the framework of the projects P20_00035 funded by FEDER/Junta de Andalucía-Consejería de Transformación Económica, Industria, Conocimiento y Universidades; LifeWatch-2019-10-UGR-01 co-funded by the Ministry of Science and Innovation through the FEDER funds from the Spanish Pluriregional Operational Program 2014–2020 (POPE) LifeWatch-ERIC action line; and PID2021-126401OB-I00, funded by MCIN/AEI/10.13039/501100011033/FEDER Una manera de hacer Europa.

**FIGURE CAPTIONS**

**Fig. 1.** The two-nested domain setup employed in the WRF experiments. In (a) the spatial coarser domain covering the Iberian Peninsula (IP) with 5 km spatial resolution is shown with the main topographical features and (b) the innermost domain focusing over Andalusia with 1 km spatial resolution, where the meteorological stations for the year 2018 are presented. Meteorological stations with pr data are represented with circles, while those that also have temperature data are represented with squares and those with only temperature data are represented with triangles. Different colors for the stations indicate the organization they are part of.

**Fig. 2.** (a) ERA5 annual precipitation anomalies expressed in mm/year for 2018 compared to 1992-2021, and (b) maximum and (c) minimum temperature anomalies (in ºC) for the same year.

**Fig. 3.** Accumulated precipitation in mm for (a) the entire year 2018 and (b) autumn (SON) for the same year, from the different observational gridded datasets (first row in the panels) and the 12 WRF sensitivity simulations. The accumulated pr values from meteorological stations are displayed with dots. The $r_G$ and $MAE_G$ values in the right bottom corner of each map indicate the pattern correlation and the mean absolute error in mm compared to AEMET-Grid. In the same way, $r_S$ and $MAE_S$ show the values compared to stations.

**Fig. 4.** As Fig. 3, but for the annual average of (a) maximum and (b) minimum temperature.

**Fig. 5.** Annual cycle of monthly (a-c) precipitation, (d-f) maximum temperature and (g-i) minimum temperature for three ranges of altitude levels above sea level: (a, d, g) lower than 400 m, (b, e, h) between 400 m and 1500 m and (c, f, i) above 1500 m.

**Fig. 6.** Non-parametric Kling-Gupta Efficiency (KGE) metric for (a) annual and (b) autumn precipitation. The values of KGE with respect to stations are shown by dots.

**Fig. 7.** As Fig. 6, but for (a) maximum and (b) minimum temperatures. The values with respect to observations are depicted in dots.

**Fig. 8.** Daily percentiles simulated by the different WRF configurations vs observational data from AEMET-Grid for (a-c) precipitation, (d-f) maximum, (g-i) and minimum temperature. The results for the different regions (LAs, MAs, and HAs) are shown in columns. Observations from stations are shown as black squares.

**Fig. 9.** Evolution of Storm Emma from (a) 28[th] February to (d) 3[rd] March according to ERA5 data. The colors in these panels represent sea level pressure, and the black lines show the geopotential height at 500 hPa. (e) Temporal evolution of hourly pr, and (f) hourly accumulated pr in the station of Alpandeire (from AEMET network) during the event.

**Fig. 10.** Accumulated precipitation during Storm Emma (from February 28[th] to March 4[th]).

**Fig. 11.** As Fig. 9, but for the extreme event occurred between (a) 20th and (b) 21st October according to ERA5 data. (c) Temporal evolution of hourly pr, and (d) hourly accumulated pr in the station of Pujerra (from SAIH-S network) during the event.

**Fig 12.** As Fig. 10 but for the event occurred from 20[th] to 21[st] October 2018.

**TABLES**

**Table 1** Summary of the twelve combinations of parameterizations used in this study. For convection, the different options for each domain are indicated as well as the acronyms used in this study.

| | Microphysics configurations | | | |
|---|---|---|---|---|
| | *THOMPSON* | | | |
| Convection (d01/d02) | G3/OFF | GF/OFF | KF/OFF | OFF/OFF |
| **Acronym** | **THOMPSON-G3** | **THOMPSON-GF** | **THOMPSON-KF** | **THOMPSON-OFF** |
| | *WRF single-moment-6-class (WSM6)* | | | |
| Convection (d01/d02) | G3/OFF | GF/OFF | KF/OFF | OFF/OFF |
| **Acronym** | **WSM6-G3** | **WSM6-GF** | **WSM6-KF** | **WSM6-OFF** |
| | *WRF single-moment-7-class (WSM7)* | | | |
| Convection (d01/d02) | G3/OFF | GF/OFF | KF/OFF | OFF/OFF |
| **Acronym** | **WSM7-G3** | **WSM7-GF** | **WSM7-KF** | **WSM7-OFF** |

**Table 2** Observational datasets description and its use in this study. With pr, tasmax, and tasmin, the precipitation, maximum and minimum temperatures are denoted.

| | Spatial resolution | Temporal resolution | Variables | Comparison method |
|---|---|---|---|---|
| CMORPH | 0.07° | 30 minutes | pr | |
| IMERG | 0.1° | 30 minutes | pr | Interpolated to AEMET-Grid |
| CHIRPS | 0.05° | daily | pr | |
| AEMET-Grid | 5 km | daily | pr, tasmax, and tasmin | |
| Stations | | hourly/daily | pr, tasmax and tasmin | Nearest point in AEMET-Grid |

**Table 3** Meteorological stations used in this study. Different sources were used: (1) the Spanish Meteorological Agency (AEMET), (2) the Autonomous Agency for National Parks (OAPNs), and the Automatic Hydrological Information System (SAIH) from (3) Hidrosur network (SAIH-S) and (4) from Guadalquivir Basin network (SAIH-G).

| Institution | Temporal resolution | Variables | stations with precipitation (temperature) |
|---|---|---|---|
| AEMET | daily/hourly | pr, tasmax and tasmin | 303 (184) |
| OAPNs | daily | | 4 (6) |
| SAIH-S | daily | pr | 119 |
| SAIH-G | daily | pr, tasmax and tasmin | 173 (29) |

**Table 4** Regions used in the study of the annual cycles of the monthly precipitation and temperature.

| | Altitude range (m) | Number of grid points | stations with precipitation (temperature) |
|---|---|---|---|
| Low altitudes (LAs) | h < 400 | 1291 | 315 (121) |
| Middle altitudes | 400 < h < 1500 | 1456 | 276 (94) |
| High altitudes (HAs) | h > 1500 | 85 | 8 (6) |

**Feliciano Solano Farias:** Formal analysis, Validation, Data Curation, investigation, Writing-Original draft preparation, Writing-Reviewing and Editing.

**Matilde García-Valdecasas Ojeda**: Conceptualization, Methodology, Writing-Original draft preparation, Writing-Reviewing and Editing, supervision.

**David Donaire-Montaño:** Software, Formal analysis, investigation, Writing-Original draft preparation, Writing-Reviewing and Editing.

**Juanjo José Rosa-Cánovas**: Investigation, visualization.

**Yolanda Castro-Díez**: Writing-Reviewing and Editing, Supervision.

**María Jesús Esteban-Parra**: Writing-Reviewing and Editing, Supervision, Funding Acquisition.

**Sonia R. Gámiz-Fortis**: Writing-Original draft preparation, Writing-Reviewing and Editing, Funding acquisition, project administration.

**Declaration of interests**

☒The authors declare that they have no known competing financial interests or personal relationships that could have appeared to influence the work reported in this paper.

☐The authors declare the following financial interests/personal relationships which may be considered as potential competing interests:

1. WRF fits better to AEMET than CHIRPS, CMORPH, and IMERG over Andalusia.

2. The convection has a greater impact on simulations than microphysics in Andalusia.

3. WRF captures well the annual cycle of pr and temperature, with regional variations.

4. WRF shows high capability in representing extreme pr events.

5. WSM7-GF is optimal for WRF in simulating Andalusia's climate.

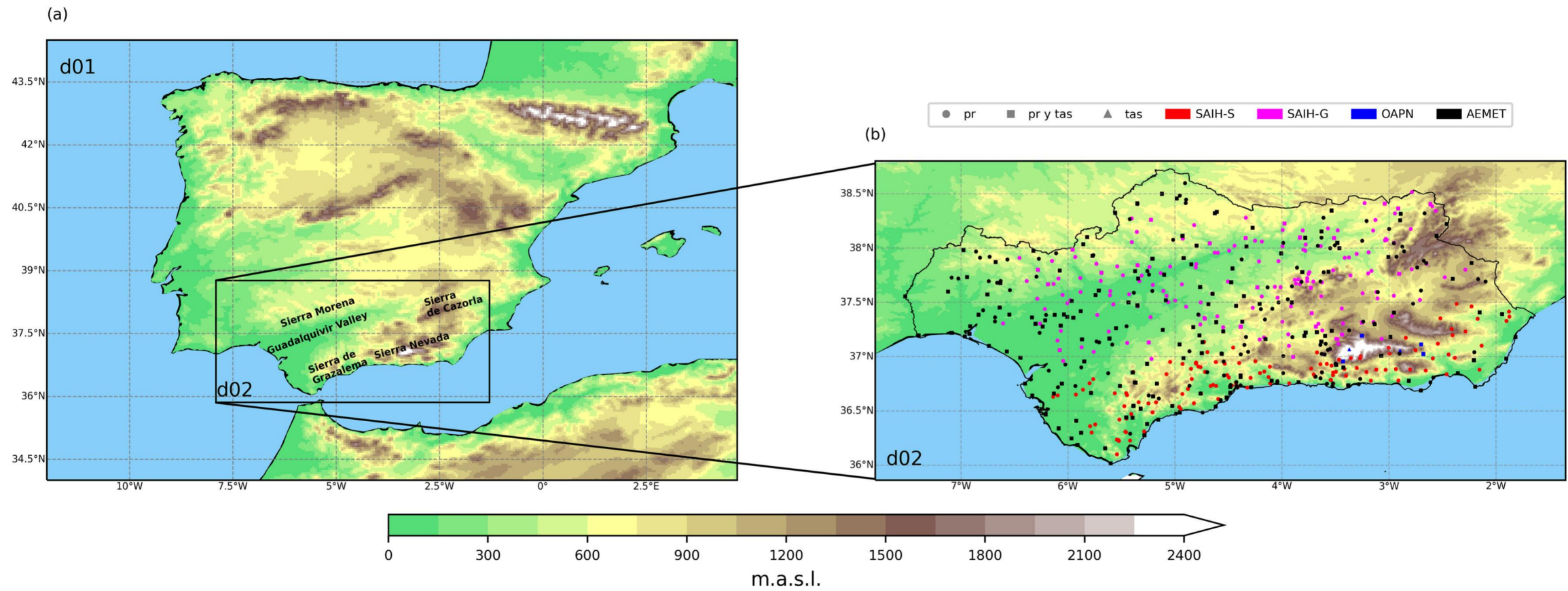

Figure 1

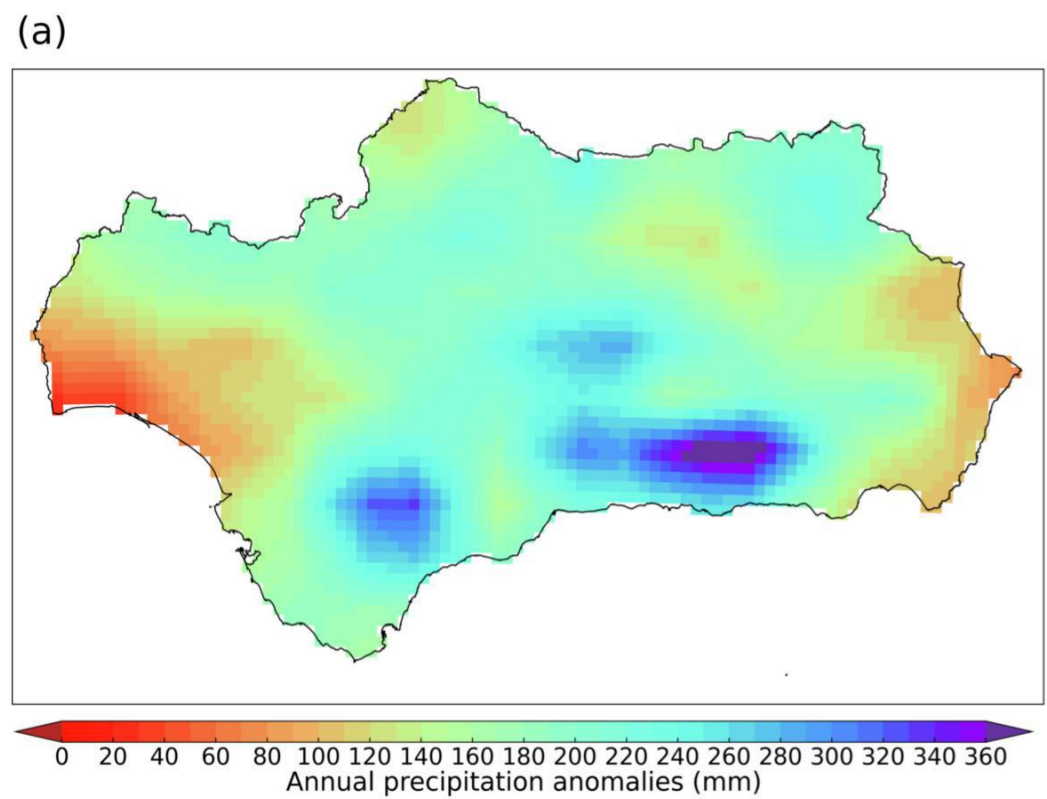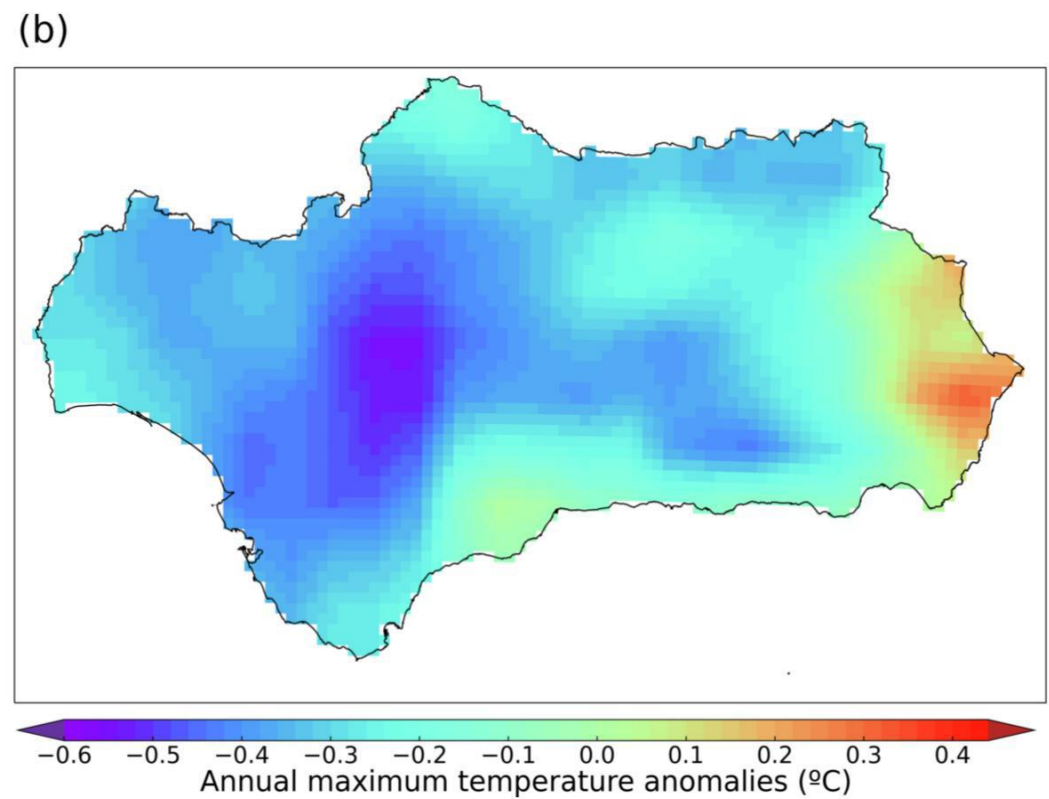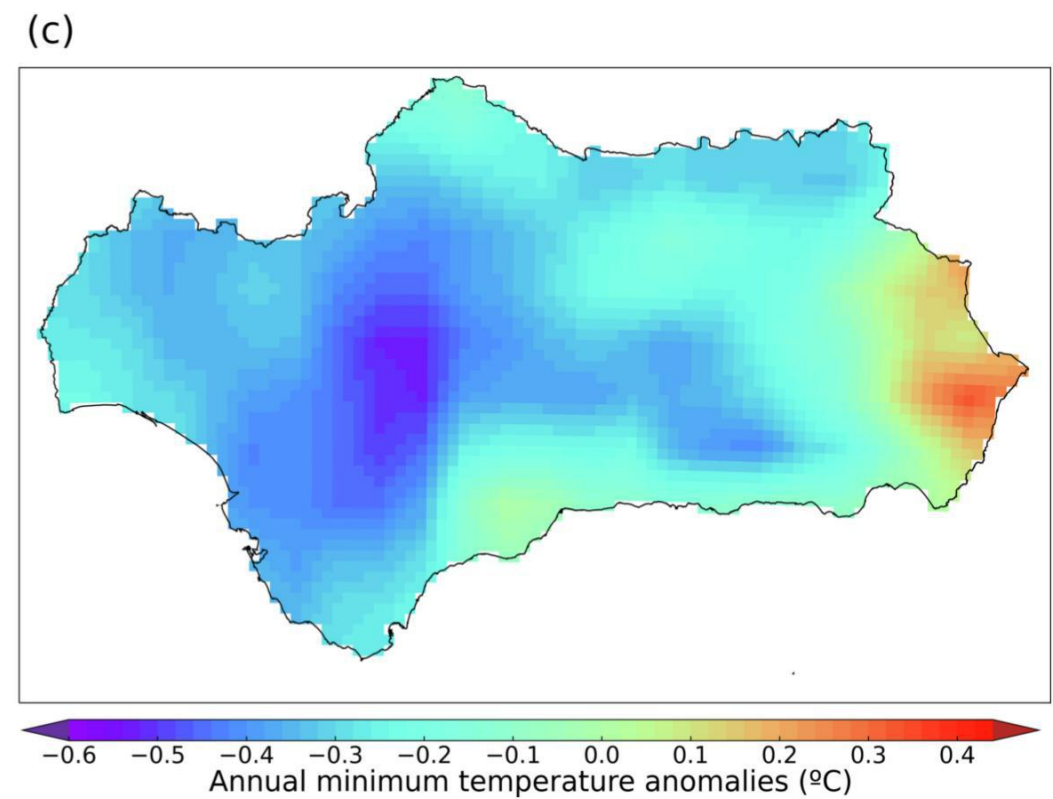

Figure 2

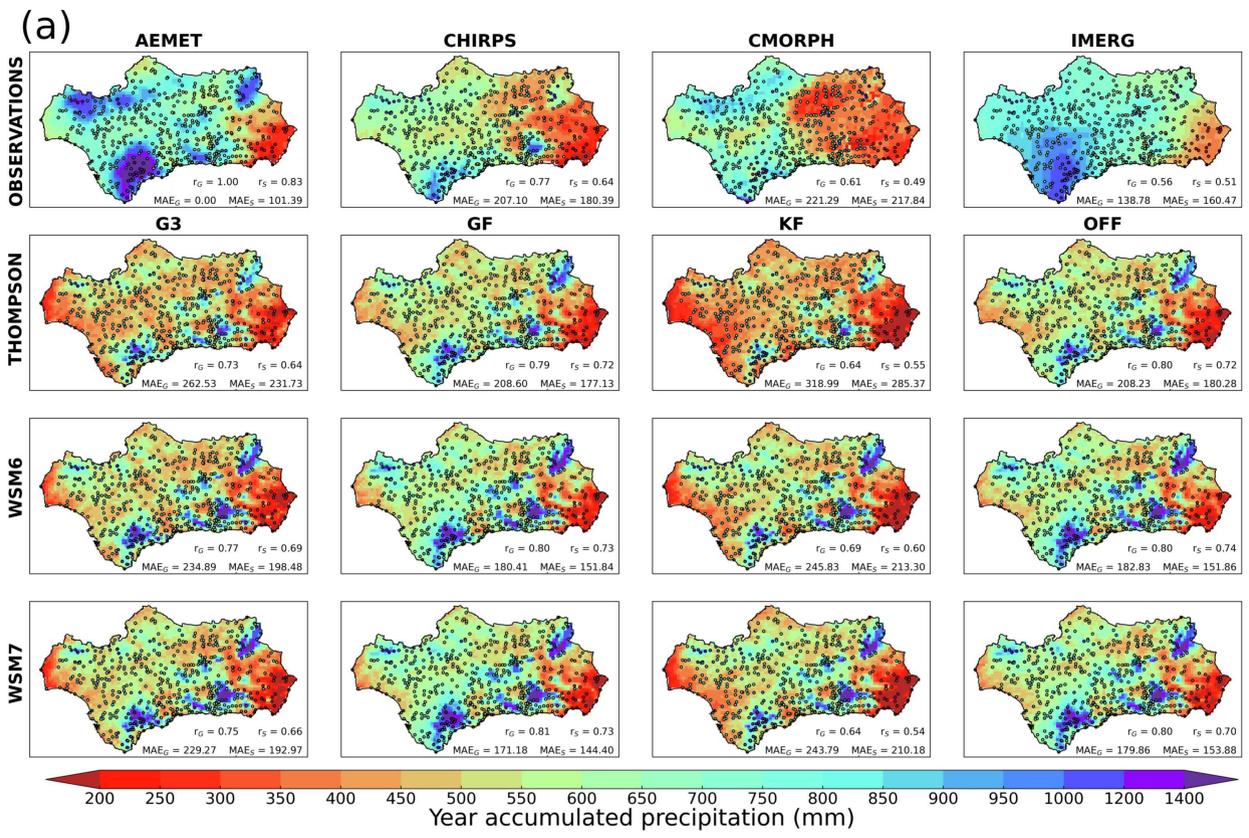

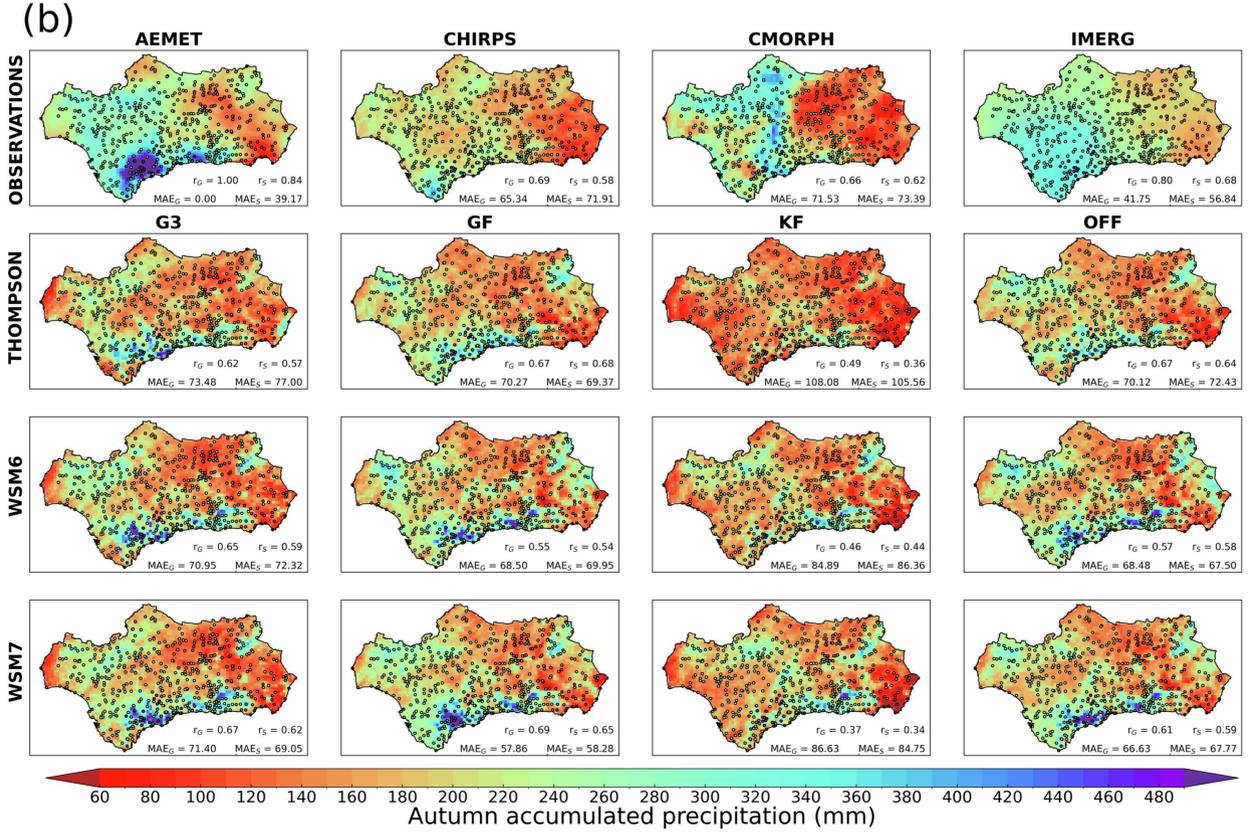

Figure 3

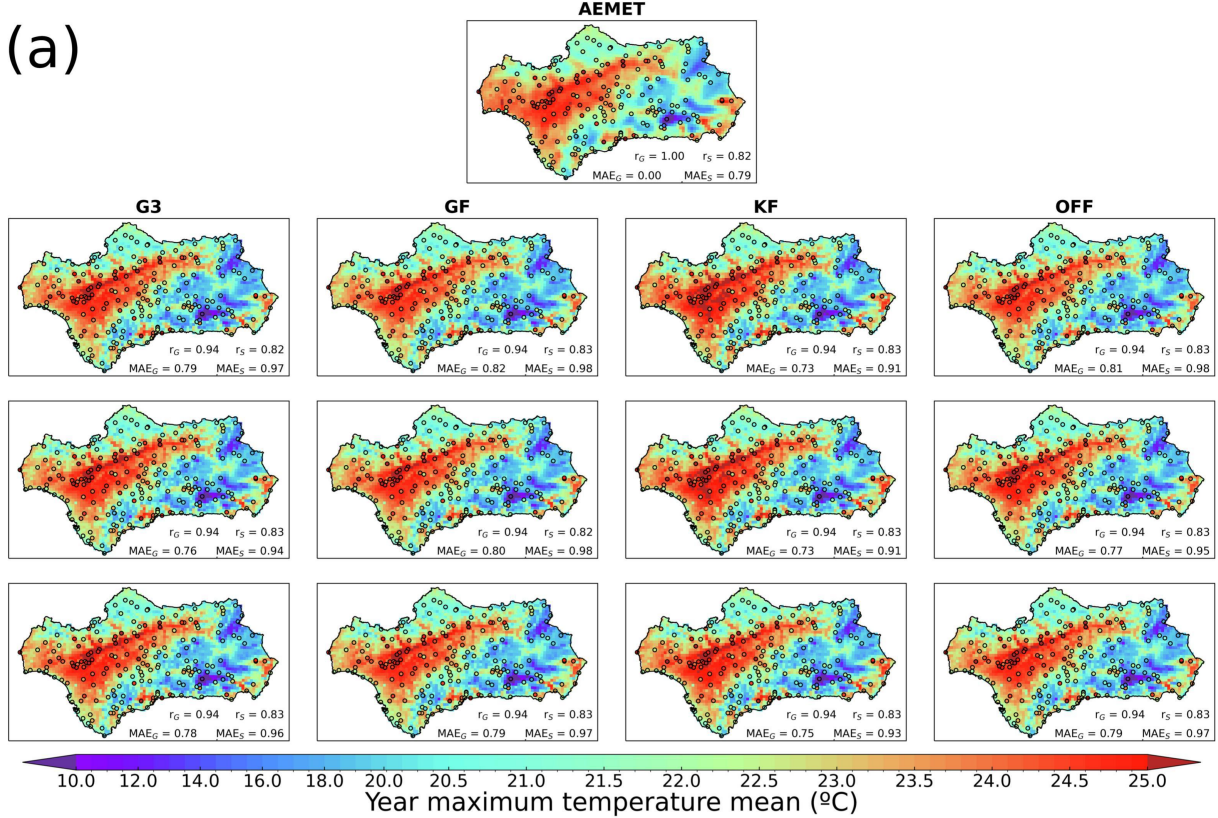

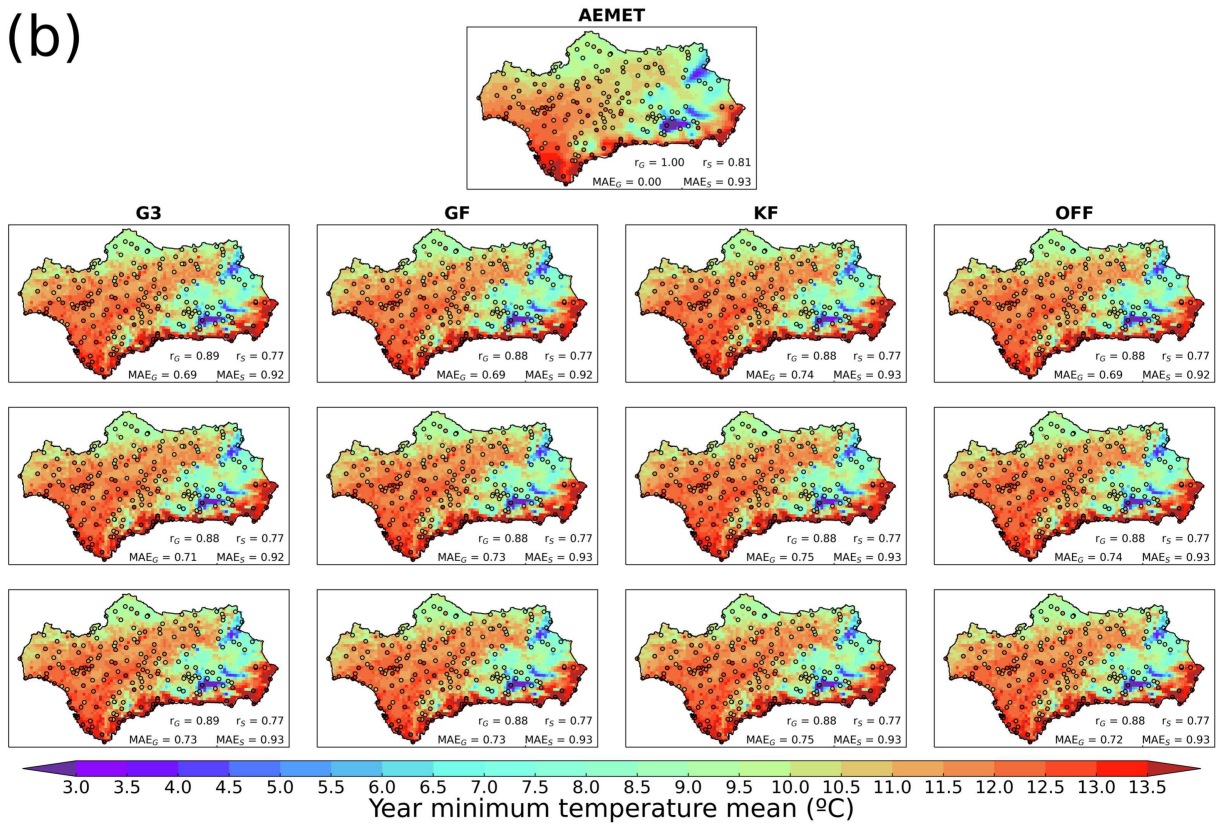

Figure 4

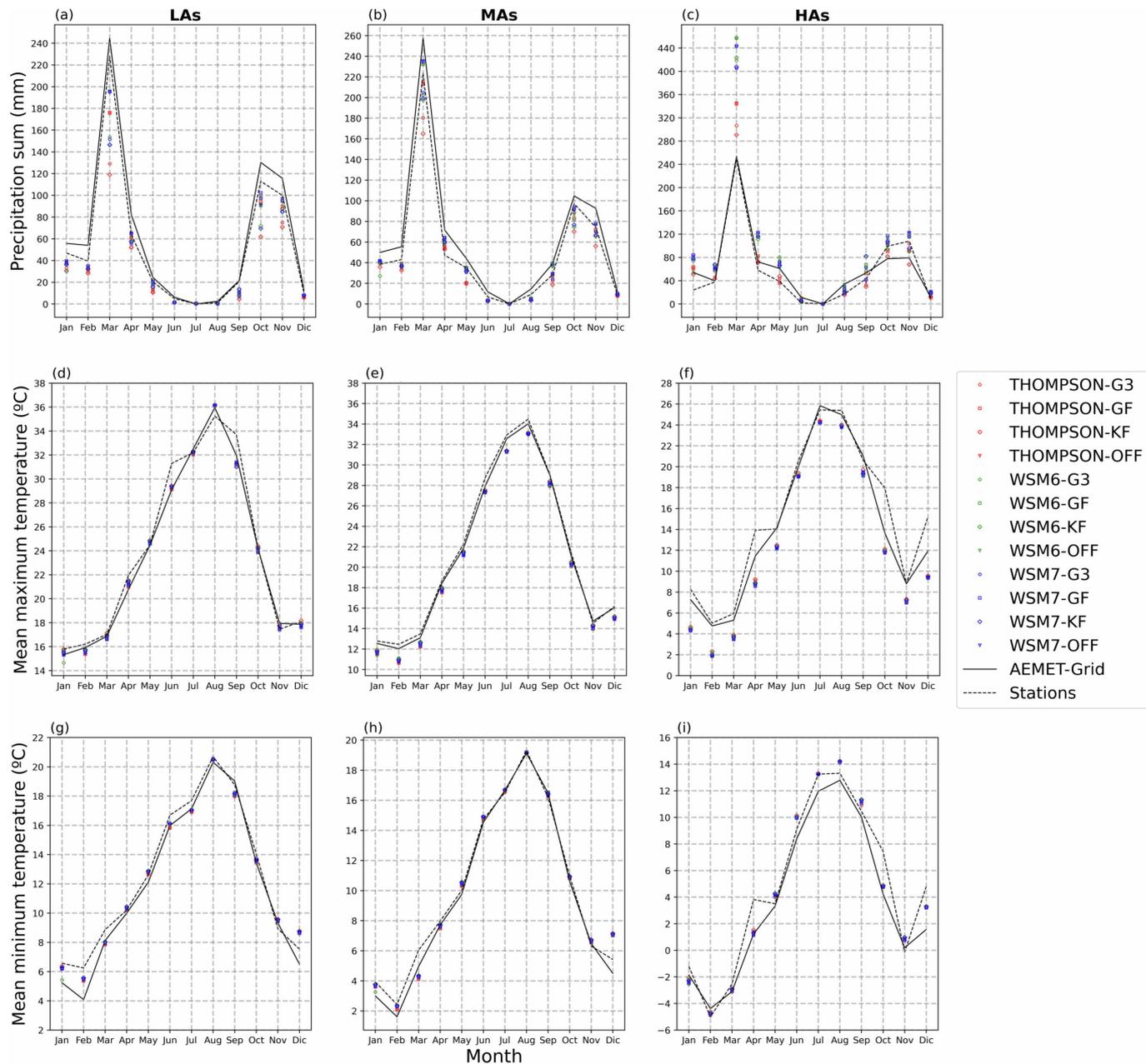

Figure 5

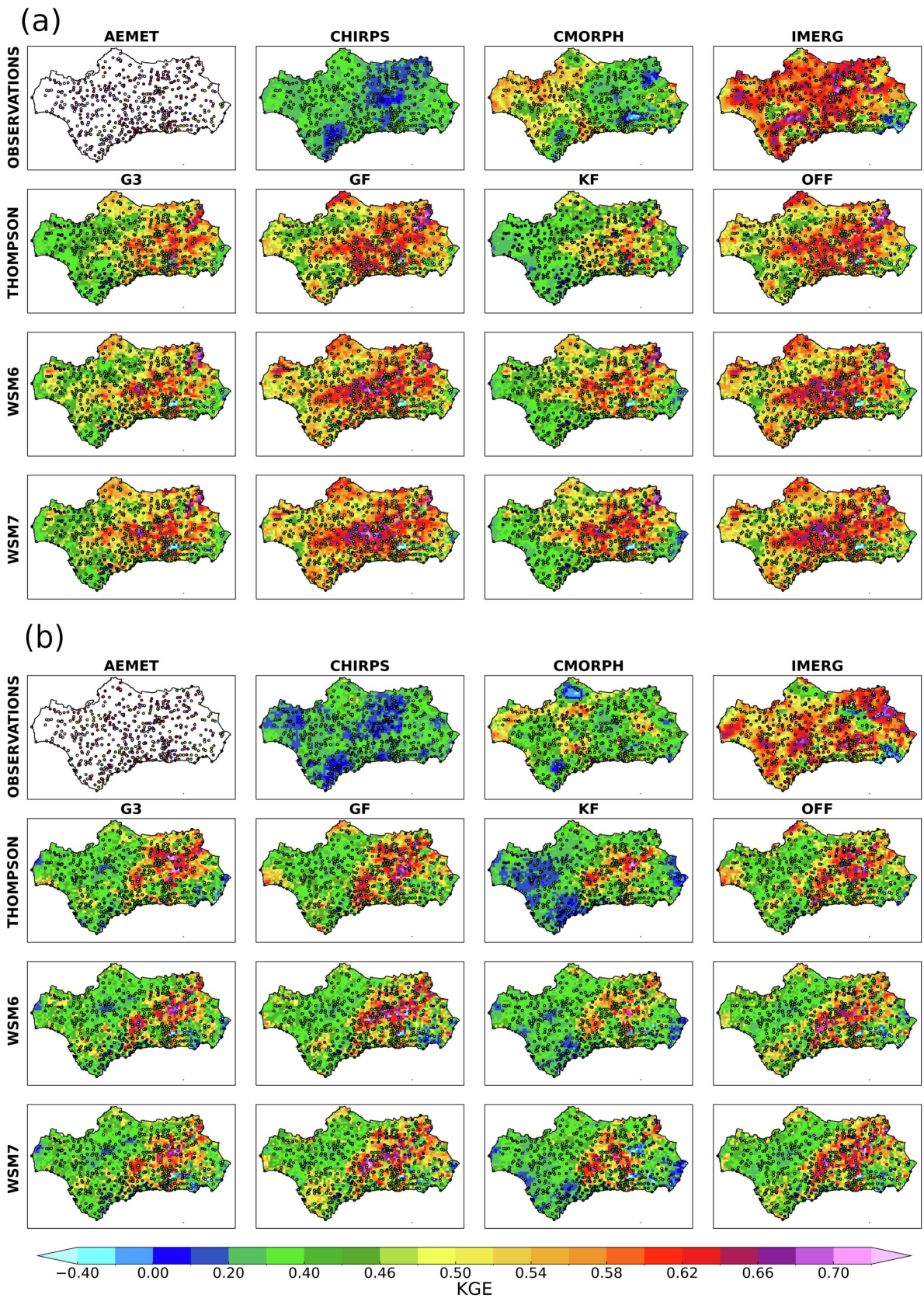

Figure 6

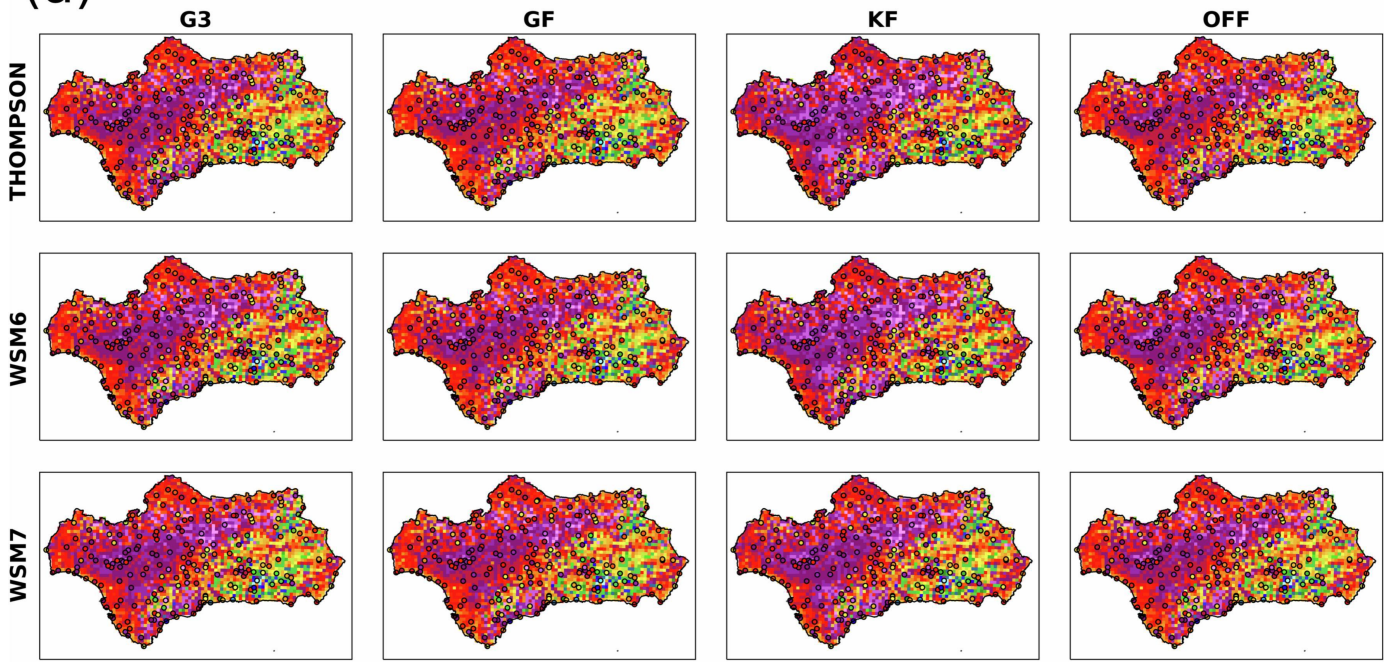

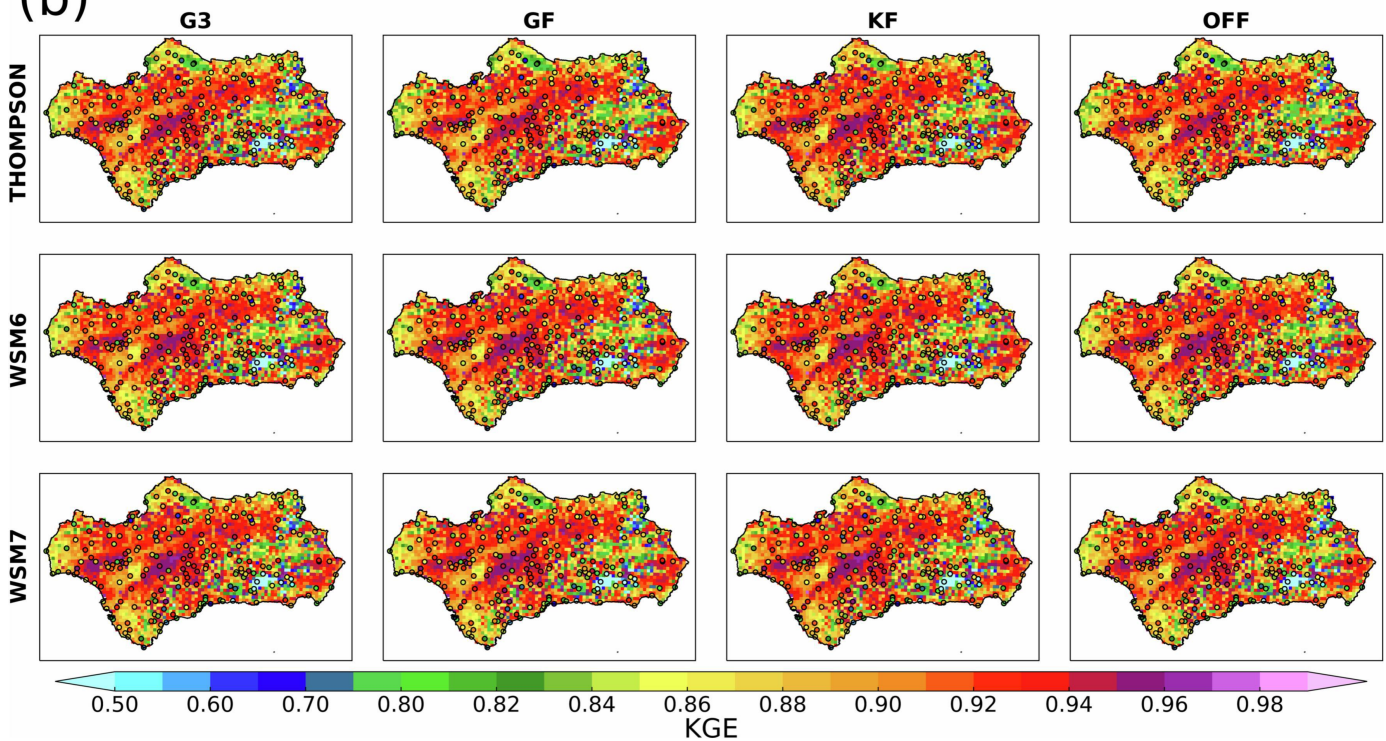

Figure 7

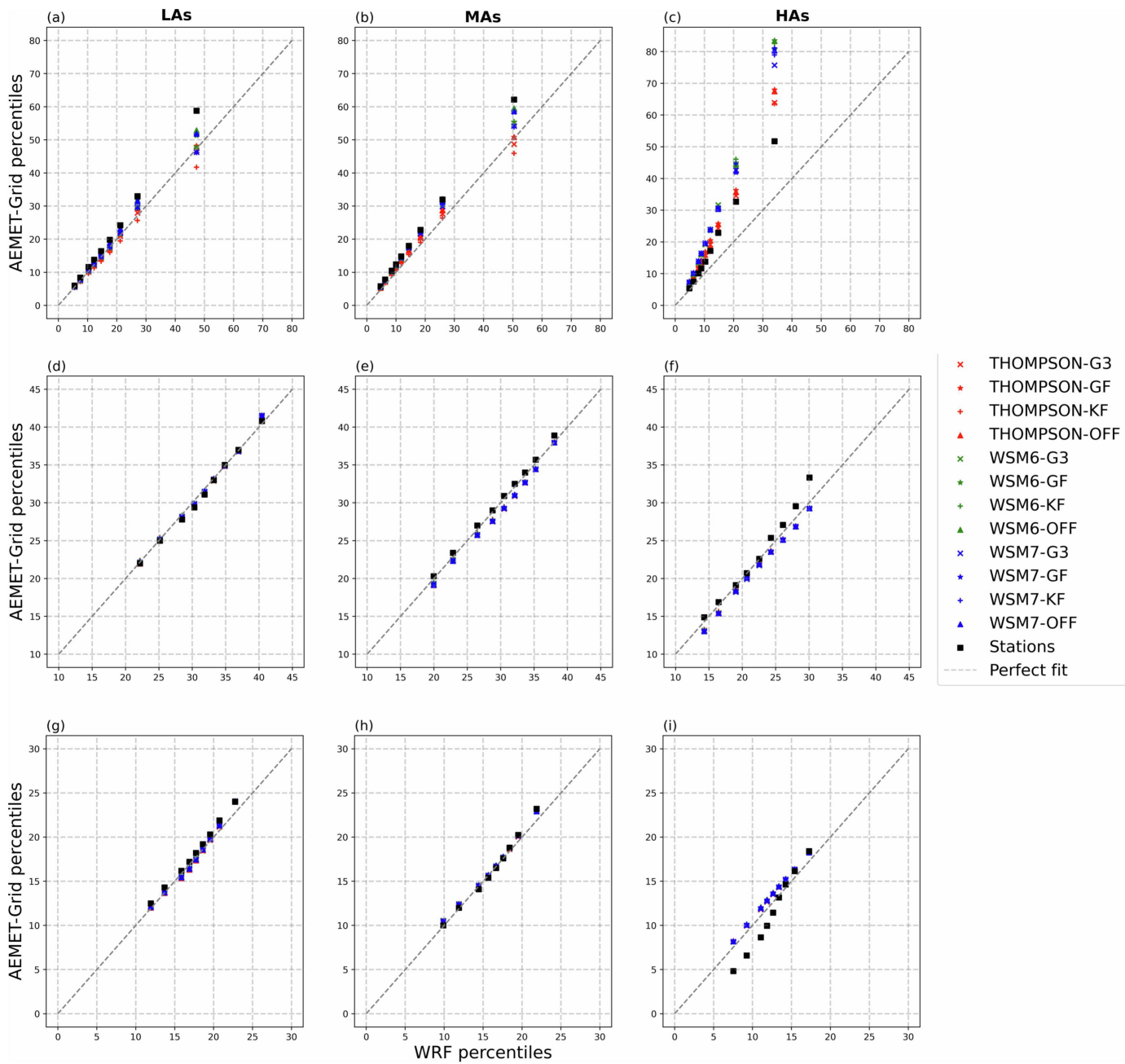

Figure 8

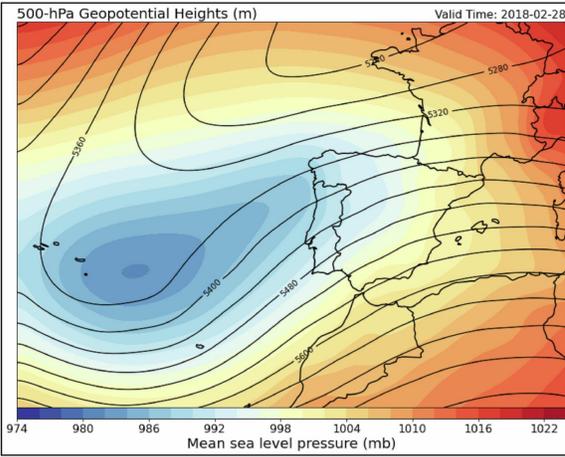

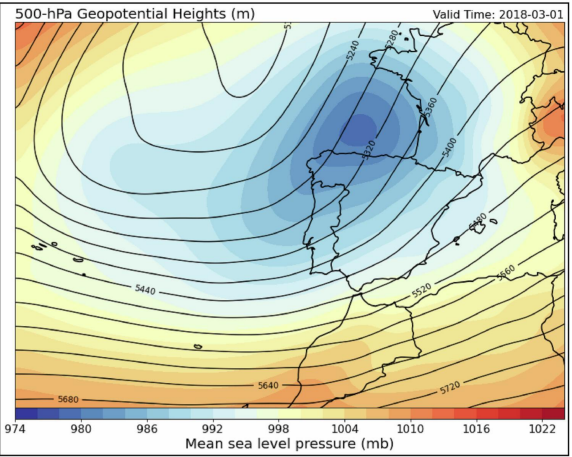

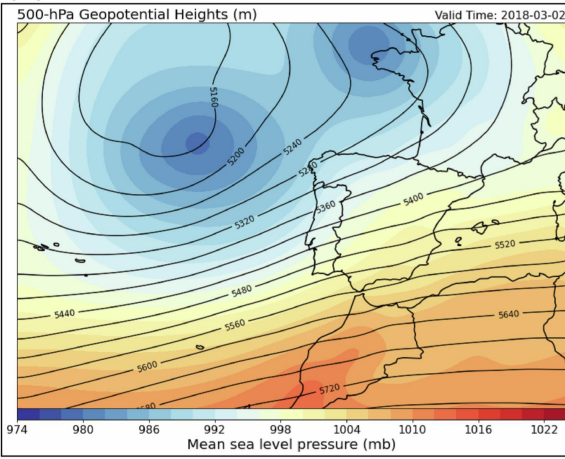

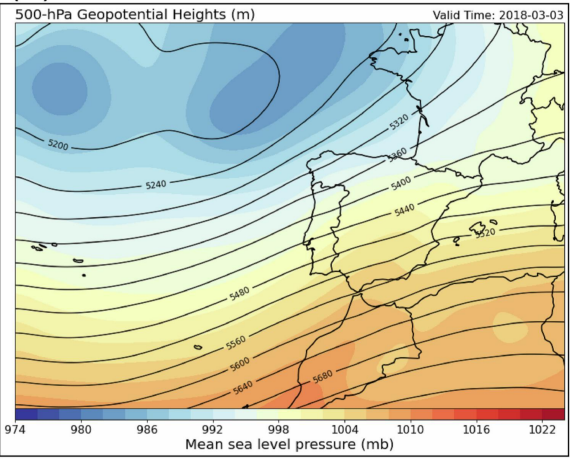

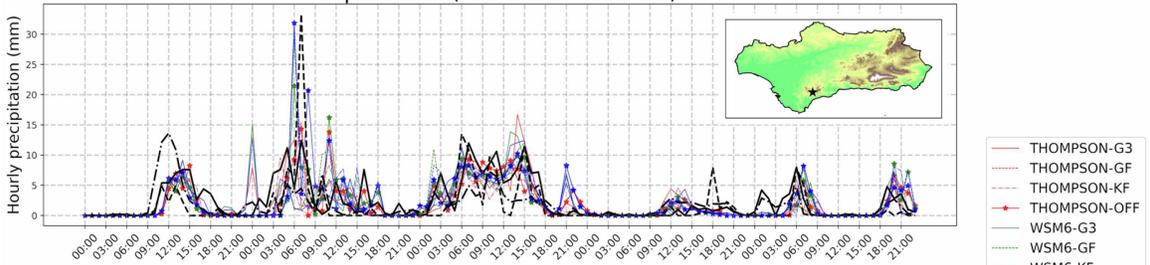

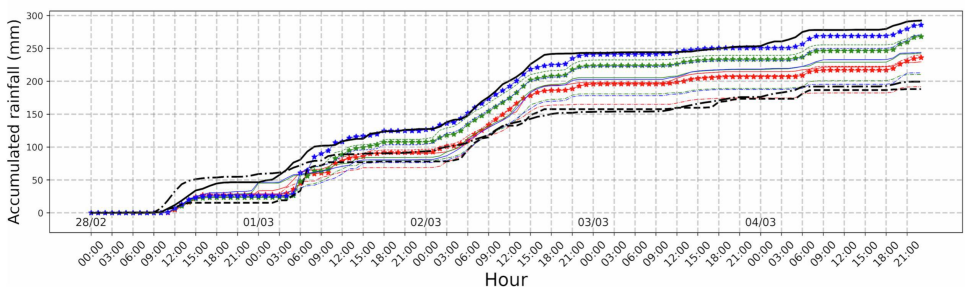

Figure 9

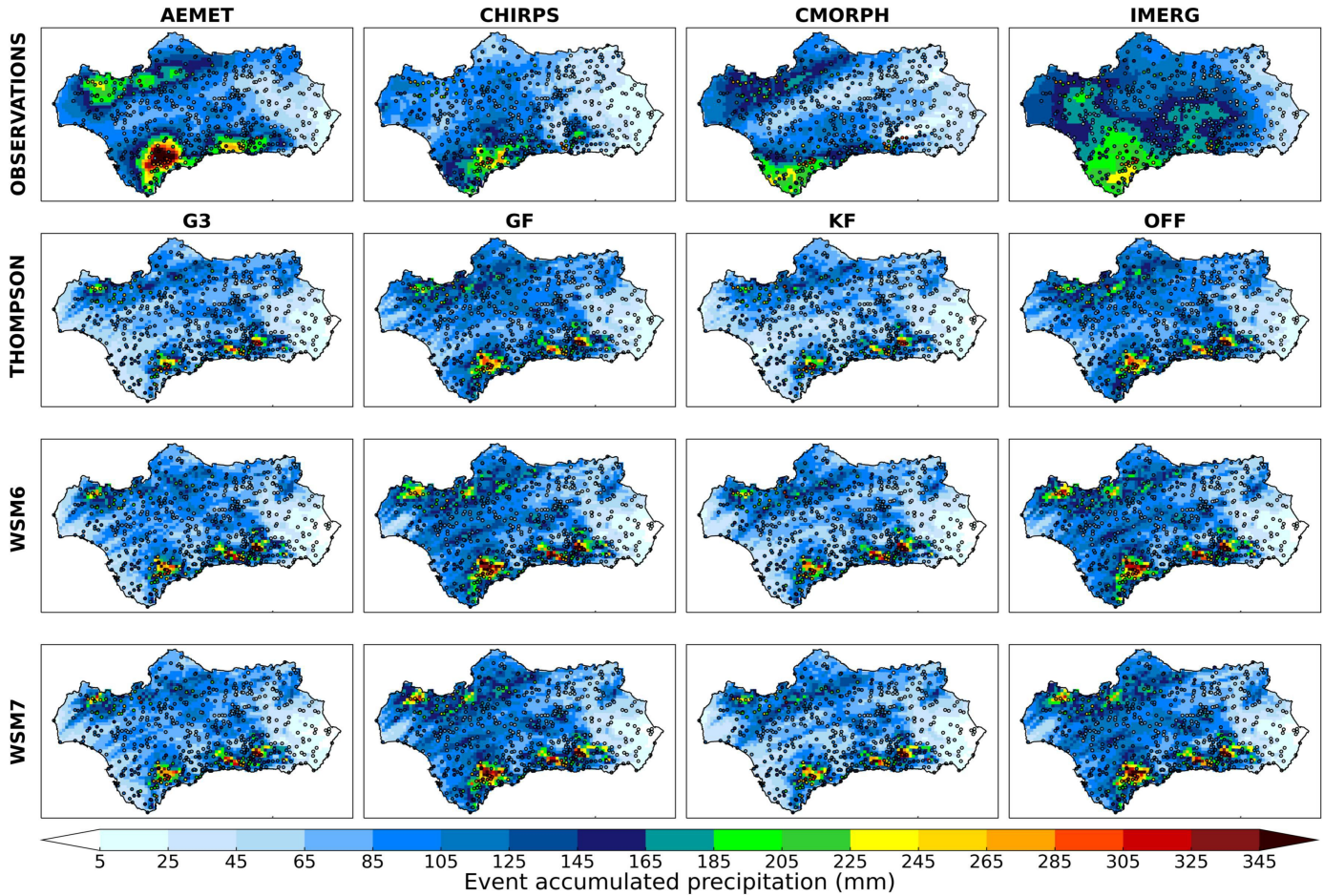

Figure 10

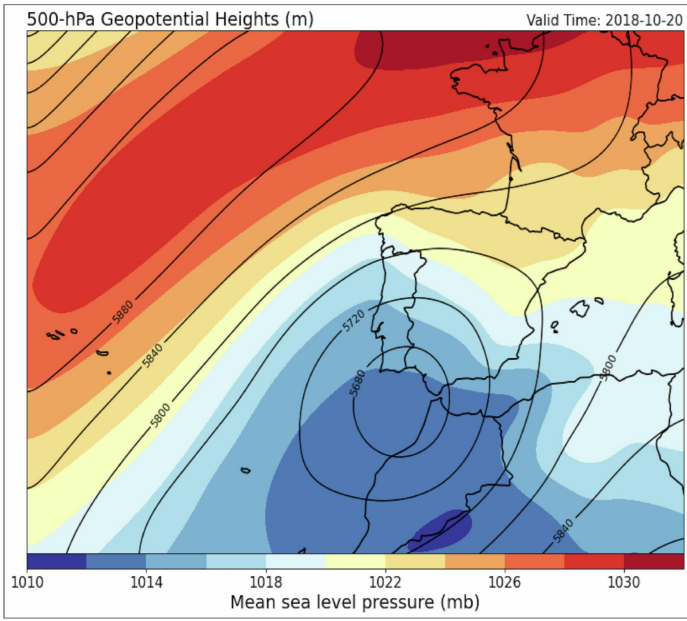
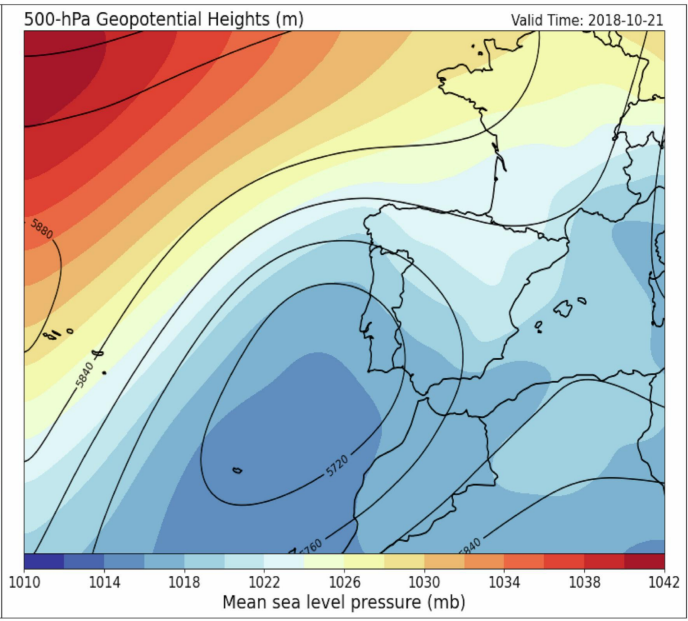
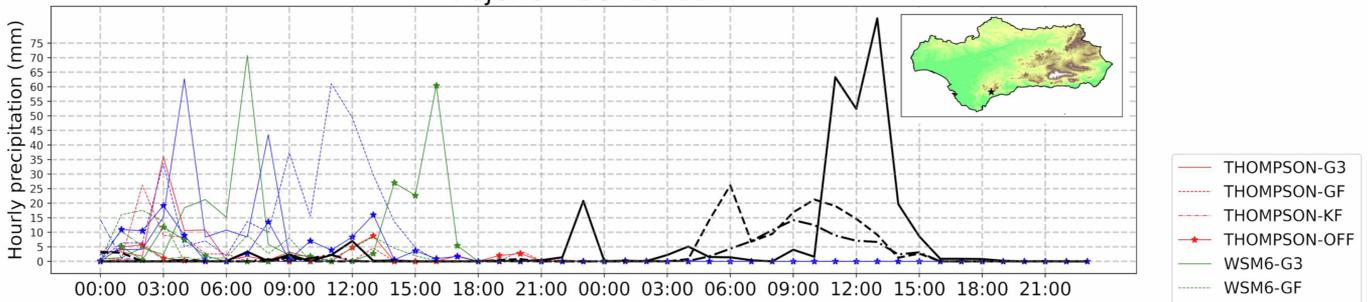
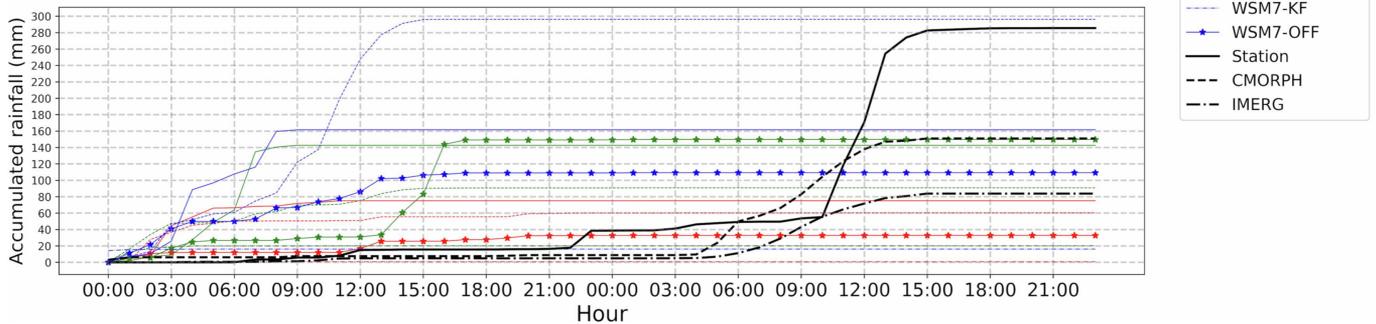

Figure 11

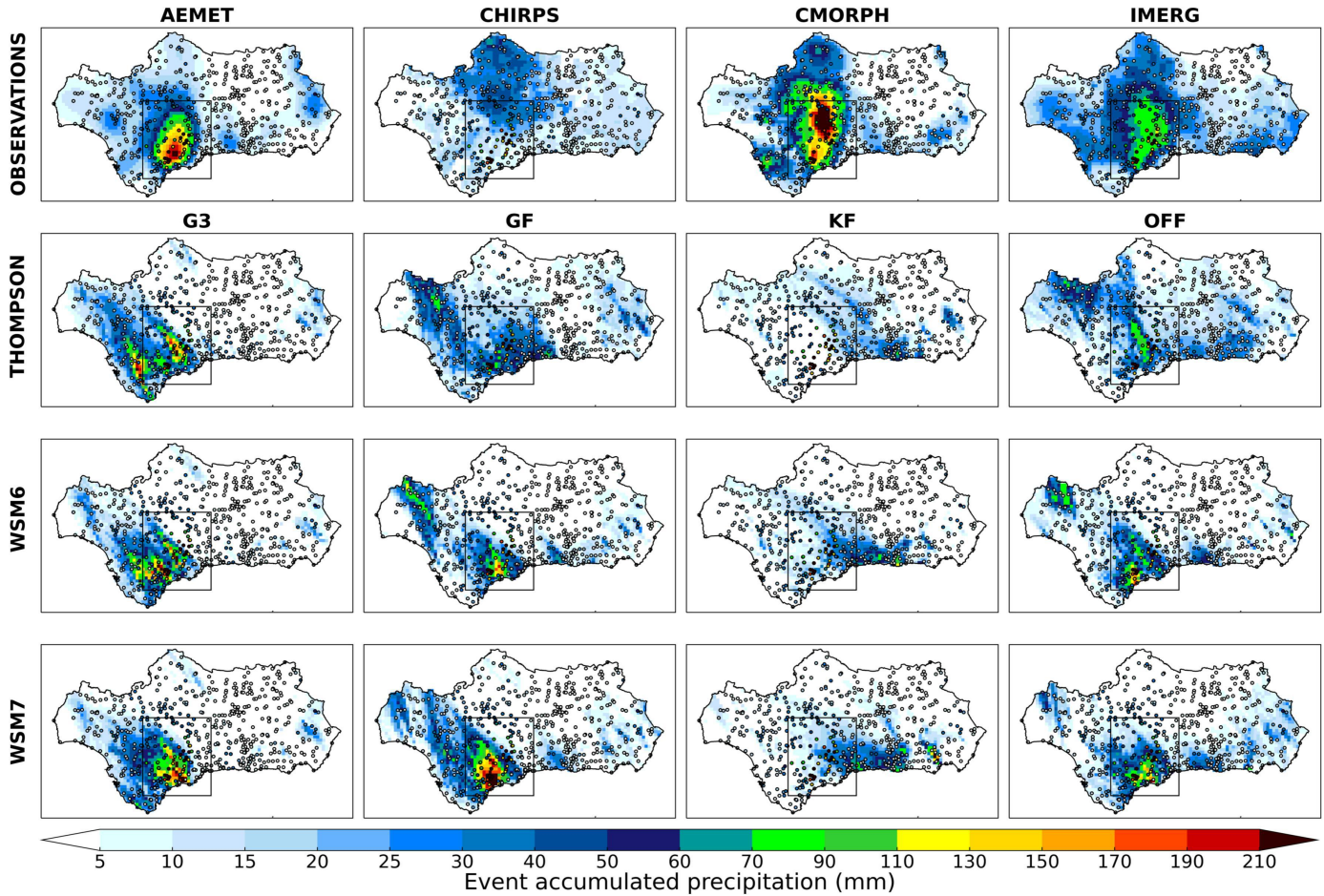

Figure 12